\documentclass[aps,prl,preprint]{revtex4}
\usepackage{CJK}

\usepackage[plainpages=false, breaklinks=true]{hyperref}
\usepackage{float}
\usepackage{graphicx}
\usepackage{amsmath}
\usepackage{caption}
\usepackage{titlesec}  
\usepackage{adjustbox}
\usepackage{setspace}
\usepackage{natbib}

\begin{document}
\begin{CJK*}{GB}{} 
	\pagenumbering{gobble}
	\title{Free energy of domain walls and order-disorder transition in a triangular lattice with anisotropic nearest-neighbor interactions}
	\author{Martina Tsvetanova}
	\affiliation{Physics of Interfaces and Nanomaterials, MESA+ Institute for Nanotechnology, University of Twente, Post Office Box 217, 7500AE Enschede, the Netherlands,\\
	Contact: m.tsvetanova@utwente.nl, h.j.w.zandvliet@utwente.nl}
	\author{Kai Sotthewes}
	\affiliation{Physics of Interfaces and Nanomaterials, MESA+ Institute for Nanotechnology, University of Twente, Post Office Box 217, 7500AE Enschede, the Netherlands,\\
	Contact: m.tsvetanova@utwente.nl, h.j.w.zandvliet@utwente.nl}
	\author{Harold J. W. Zandvliet}
	\affiliation{Physics of Interfaces and Nanomaterials, MESA+ Institute for Nanotechnology, University of Twente, Post Office Box 217, 7500AE Enschede, the Netherlands,\\
	Contact: m.tsvetanova@utwente.nl, h.j.w.zandvliet@utwente.nl}
	\maketitle
\end{CJK*}

\setlength{\abovecaptionskip}{0.5pt}
\setlength{\belowcaptionskip}{0.1pt}
\setlength{\belowdisplayskip}{0pt} 
\setlength{\abovedisplayskip}{-0pt} 

\renewcommand{\theequation}{\arabic{equation}}
\renewcommand{\thefigure}{\arabic{figure}}
\renewcommand{\thetable}{\arabic{table}}
\renewcommand{\thesection}{\arabic{section}} 
\renewcommand{\thesubsection}{\arabic{section}.\arabic{subsection}} 

\captionsetup[figure]{labelfont=bf}
\captionsetup[subfigure]{labelfont=bf}
\captionsetup[table]{labelfont=bf}

\newpage
\section{Abstract}
\pagenumbering{arabic}

We have derived exact expressions for the domain wall free energy along the three high-symmetry directions of a triangular lattice with anisotropic nearest-neighbor interactions. The triangular lattice undergoes an order-disorder phase transition at a temperature $T_c$ given by $e^{-(\epsilon_1+\epsilon_2)/2kT_c}+ e^{-(\epsilon_2+\epsilon_3)/2kT_c}+ e^{-(\epsilon_3+\epsilon_1)/2kT_c}=1$, where $\epsilon_1$, $\epsilon_2$, $\epsilon_3$ are the nearest-neighbor interaction energies and $\epsilon_1+\epsilon_2>0$, $\epsilon_2+\epsilon_3>0$, $\epsilon_3+\epsilon_1>0$. Finally, we have derived expressions for the thermally induced meandering of the domain walls at temperatures below the phase transition temperature. We show how these expressions can be used to extract the interaction energies of two-dimensional systems with a triangular lattice.

\section{Introduction}

The domain wall free energy is an important quantity, governing many phenomena \cite{Bartelt1991, Williams1994, Jeong1999}. For instance, it plays a role in order-disorder phase transitions, domain wall meandering, and the equilibrium shape of a domain embedded in a host of another phase. Here we focus on the domain wall free energy of a triangular lattice with anisotropic nearest-neighbor interactions.

The creation of a domain boundary costs energy and therefore a boundary will be perfectly straight at $T= 0$ K. The formation of kinks in a domain increases the length of the domain wall. This costs energy, but also increases the entropy. The total free energy of a domain wall, $F=U-TS$, decreases with increasing temperature.
At higher temperatures the entropy term becomes more and more significant which can lead to an order-disorder phase transition. The temperature at which the free energy of a boundary between two domains becomes zero is the critical temperature $T_c$ of the order-disorder transition. Above that temperature, kinks are generated spontaneously at the domain wall. As a consequence, if a domain of a material is embedded in a host phase, the domain wall will vanish as it will not be possible to define a boundary between the two phases.

The computation of the partition functions of crystal lattices within the framework of the solid-on-solid model and the Ising model are closely related. This is not strange as in the first case we are interested in the interaction energies between neighbor atoms which represent the bond strengths, while in the second case the framework is analogous, but the focus is on the interaction of neighboring spins. That is why, it is also expected to obtain similar expressions for the domain wall free energies, sometimes referred to as the boundary tension. For instance, Kramers and Wannier \cite{KramersWannier1941} showed that obtaining the full partition function of a 2D crystal with Ising interaction is reduced to an eigenvalue problem, which was later solved in the 2D rectangular lattice case by Onsager \cite{Onsager1944} for interactions between nearest neighbors only. Later on Kaufman proposed a spinor analysis formalism \cite{Kaufman1949}.

There is an alternative, rather straightforward route to obtain the order-disorder temperature of a two-dimensional system with nearest-neighbor interactions. By deriving an expression for the domain wall free energy and setting this expression equal to zero, one finds the critical temperature of a system and reduces the problem to a quasi 1D case. This method provides the exact solution for the square lattice with isotropic nearest-neighbor interactions and for arbitrarily oriented step edges on a square lattice with anisotropic nearest-neighbor interactions \cite{Zandvliet2000, Zandvliet2015}. Later on, next-nearest-neighbor interactions were also introduced in the model \cite{Zandvliet2006}. Within the solid-on-solid framework, it was demonstrated that expressions for the step free energies along the high-symmetry directions of a square lattice are identical to the exact expressions for the boundary tension of the two-dimensional square lattice with Ising interaction, derived by Onsager \cite{Zandvliet2000, Zandvliet2015, Onsager1944}.

Apart from the square lattice, attention has also been devoted to the triangular lattice with isotropic nearest-neighbor interactions. Expressions for the partition sum, the free energy, and the mean squared kink length were already derived for the domain wall of a triangular lattice with isotropic nearest-neighbor interactions, which also allows to extract the critical temperature of the order-disorder transition \cite{Kai2013a, Kai2013}. In this paper, with the use of statistical mechanics considerations in the solid-on-solid model, we obtain the partition sum for the boundaries of the 2D triangular lattice with anisotropic nearest-neighbor bond strengths, by deploying the quasi 1D method, put forward in detail in ref. \citenum{Zandvliet2015}. We pay attention to the order-disorder transition in the lattice and compare our conditions for the order-disorder transition to the conditions obtained for the triangular Ising model with anisotropic interactions, derived by Houtappel \cite{Houtappel1950}. Our method requires less computation compared to the approach of Houtappel, and differently from ref. \citenum{Houtappel1950}, we provide exact expressions for the boundary tension (the domain wall free energy). Additionally, by using our exact expression for the partition function, we obtain an expression for the mean squared kink length of a domain wall and demonstrate how it can be applied to experimental data in order to extract the interaction energies.

\section{Partition sum}

In order to determine the domain wall free energy, we first need to obtain the partition sum for the domain wall of the 2D triangular lattice. For clarity, we introduce the simple counting method illustrated in Figure \ref{fig:kink diagram}. In Figure \ref{fig:kink diagram} (A) the center of the illustrated honeycomb lattice corresponds to a lattice point of the triangular lattice (atoms/lattice positions are displayed as gray circles, the unit cell with black dashed lines). Please note that the honeycomb grid is the dual lattice of the triangular lattice \cite{Baxter1982}. The three high-symmetry directions of the triangular lattice are denoted as well. Kinks can extend in the positive (upwards) or the negative (downwards) direction, as shown in Figure \ref{fig:kink diagram} (B). Each kink corresponds to a unit cell being added or removed from the boundary. The probability of the occurrence of a certain kink is given by Boltzmann statistics, resulting in an exponential term $e^{-E/kT}$ in the partition sum, where $E$ is the kink formation energy, $k$ is the Boltzmann constant, and $T$ is the temperature. The problem of obtaining the partition function is, therefore, reduced to the counting of the kinks in a systematic fashion. To do so, we make use of the dual lattice system: we draw the boundaries between atoms and the resulting lattice of line segments is a honeycomb lattice. Each line segment of the honeycomb lattice corresponds to a present/broken bond. The honeycomb line segments are perpendicular to the bonds between atoms, illustrated on the right in Figure \ref{fig:kink diagram} (A). Thus, if the interaction energies perpendicular to the three high-symmetry directions are respectively  $\epsilon_1$, $\epsilon_2$, and $\epsilon_3$, then the honeycomb line segments along the high-symmetry directions will correspond to energies $\epsilon_1/2$, $\epsilon_2/2$, and $\epsilon_3/2$, respectively. At the boundary itself, each honeycomb segment represents a broken bond. The kink formation energy corresponds to the sum of the line segments values which participate in the kink.

We express the kink formation partition sum for all positive and negative kinks. The total partition sum that follows is: 

\begin{align}\label{eq:z_sum}
Z= \sum_i e^{-E_i/kT}= e^{-\epsilon_3/2kT}\Big[ \sum_{n=0}^{\infty} e^{-(n\epsilon_1+(n+1)\epsilon_2)/2kT}+ e^{-(n\epsilon_2+(n+1)\epsilon_1)/2kT}\Big]
\end{align}
\smallskip

\noindent After simplifying Equation (\ref{eq:z_sum}), the partition sum becomes (see the Appendix for a detailed derivation):

\begin{align} \label{eq:Z}
Z= \frac{e^{-\epsilon_3/2kT}(e^{-\epsilon_1/2kT}+e^{-\epsilon_2/2kT})}{1-e^{-(\epsilon_1+\epsilon_2)/2kT}}
\end{align}
\smallskip

Note that Equation (\ref{eq:Z}) was obtained considering the high-symmetry directions orientation as on the diagram in Figure \ref{fig:kink diagram} (A). To obtain the partition sums along the remaining two high-symmetry configurations (at $60^{\circ}$ with respect to the current arrangement), it is needed to permute the indices of the interaction energies. Also, so far we only made use of the honeycomb lattice for accounting purposes; the derived partition function is applicable to the triangular lattice and not to the honeycomb lattice.

\section{Free energy and equilibrium shape}

To derive expressions for the free energy, the well-known equation from statistical mechanics is used, or: $F= -kT\ln(Z)$. This leads to the following expressions for the domain wall free energy along the three high-symmetry directions, given the condition $\epsilon_1>\epsilon_2>\epsilon_3$ without a loss of generality (see the Appendix for a step-by-step derivation):

\begin{align}\label{eq:free energy 1}
F_{\hat{\epsilon_3}}= F_{32}=\frac{\epsilon_3+\epsilon_2}{2} - kT\Big[ \ln{(1+ e^{(\epsilon_2-\epsilon_1)/2kT})} -  \ln{(1- e^{-(\epsilon_1+\epsilon_2)/2kT})} \Big]
\end{align}

\begin{align}\label{eq:free energy 2}
F_{\hat{\epsilon_1}}= F_{13}=\frac{\epsilon_1+\epsilon_3}{2} - kT\Big[ \ln{(1+ e^{(\epsilon_3-\epsilon_2)/2kT})} -  \ln{(1- e^{-(\epsilon_2+\epsilon_3)/2kT})} \Big]
\end{align}

\begin{align}\label{eq:free energy 3}
F_{\hat{\epsilon_2}}=F_{23}=\frac{\epsilon_2+\epsilon_3}{2} - kT\Big[ \ln{(1+ e^{(\epsilon_3-\epsilon_1)/2kT})} -  \ln{(1- e^{-(\epsilon_3+\epsilon_1)/2kT})} \Big]
\end{align}
\smallskip

The free energy for the three high-symmetry directions is plotted in Figure \ref{fig:free energy plot} per unit length of the domain wall. Both on the plot and in equations (\ref{eq:free energy 1})-(\ref{eq:free energy 3}) we notice that for two out of the three high-symmetry directions, the free energies at 0 K are equal. Furthermore, all of the free energies per unit length of the domain wall at 0 K correspond exactly to the sum of two broken bonds. The physical explanation follows from the fact that the system always prefers the lowest possible free energy. The possible domain wall arrangements following from the current analysis are shown in Figure \ref{fig:boundary diagram}. For the $\hat{\epsilon_2}$ and $\hat{\epsilon_3}$ high-symmetry directions the domain walls at 0 K are indistinguishable and have a lower energy $(\epsilon_2/3+\epsilon_3/2)$ per unit length than the domain wall corresponding to the $\hat{\epsilon_1}$ high-symmetry direction which has an energy of $(\epsilon_1/2+\epsilon_3/2)$ per unit length. Of course, all these boundary formation energies should be larger than 0. A negative boundary energy implies that the boundary cannot exist.

We note that the domain walls described with the segments in Figure \ref{fig:boundary diagram} deviate from the orientation of the high-symmetry directions shown in Figure \ref{fig:kink diagram} (A). The outcome in Figure \ref{fig:boundary diagram} suggests that at 0 K, the unkinked boundaries are composed only out of consecutive 0 kinks (for a more detained discussion regarding the kinks see the mean squared kink length discussion in the Appendix). Thus, to have orientation along either $\hat{\epsilon_1}$, $\hat{\epsilon_2}$, or $\hat{\epsilon_3}$ as defined in Figure \ref{fig:kink diagram} (A), the domain walls must have experienced thermal meandering. Note that, still there is no loss of mathematical generality in our model as the kink accounting system would remain the same because it is irrespective from the choice of the boundary, as long as the broken bonds are correctly counted. Additionally, the simplistic sketch of a triangular island presented in Figure \ref{fig:kink diagram} (A) will not hold, as boundaries that have higher free energy per unit length will be expected to be shorter (in the current example domain walls along $\hat{\epsilon_1}$) than domain walls that cost less energetically (here these are the boundaries along $\hat{\epsilon_2}$ and $\hat{\epsilon_3}$). We will focus on the thermal domain wall meandering later in this article.

\begin{figure}[ht]
	\centering
	\includegraphics[width=0.5\linewidth]{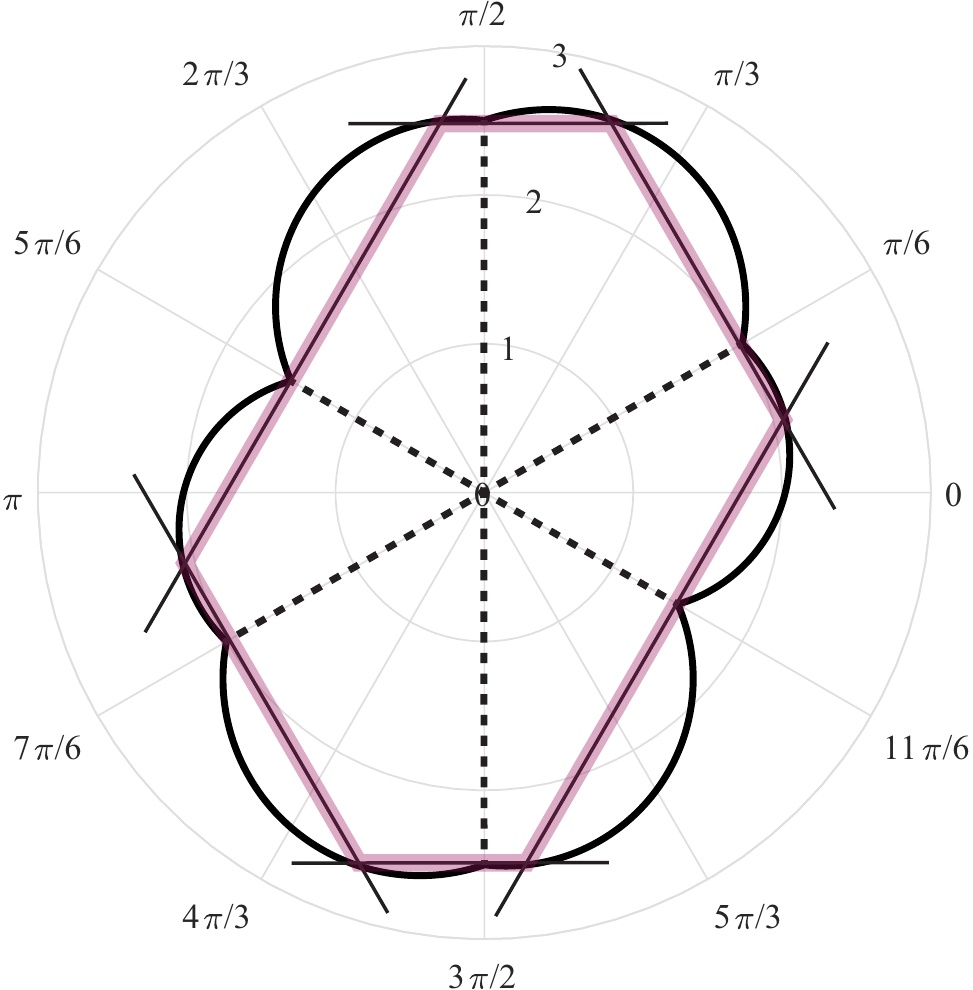}
	\smallskip
	\caption{\small Equilibrium island shape at 0 K for the triangular lattice with anisotropic nearest-neighbor interactions. The polar plot of the energy is shown with a thick black line. The Wulff construction is made with the use of the black dashed lines and thinner black lines. The resulting equilibrium shape is highlighted with a transparent (purple) overlay. The shape is constructed for $\epsilon_1>\epsilon_2>\epsilon_3$, or $\epsilon_1= 3\epsilon$, $\epsilon_2=2\epsilon$, $\epsilon_3=\epsilon$ for an arbitrary $\epsilon$. How the polar plot was obtained is explained in the Appendix.}
	\label{fig:equil_shape}
\end{figure}

So far, here we presented an analysis of the free energy for freely meandering domain walls. We extend the current analysis and also provide the equilibrium island shape for the triangular lattice with anisotropic nearest-neighbor interactions. We restrict ourselves to $T= 0$ K as at 0 K there is no entropy contribution and the shape can be obtained as explained in the corresponding Appendix section. At higher temperatures, the method presented in the Appendix needs to be extended to incorporate the contribution from the entropy term in the expression for the free energy. To do so, one must study the domain walls along directions, different from the principal directions discussed in this paper (domain walls with arbitrary orientation). The equilibrium island shape here is obtained in two steps: first a polar plot of the domain wall energy per unit length is obtained and afterwards the Wulff construction is made. The result is presented in Figure \ref{fig:equil_shape}. Note the distorted hexagonal shape which results due to the unequal nearest-neighbor interactions. We observe in the case of $\epsilon_1= 3\epsilon$, $\epsilon_2=2\epsilon$, and $\epsilon_3=\epsilon$ (for an arbitrary $\epsilon$), three unequal side lengths. The ratio 1:2:3 is kept; the boundary free energy cost is inversely proportional to the domain wall length. Although the shape at higher temperature is not provided, the one at 0 K is already instructive for what can be expected. At temperatures higher than 0 K, the shape will deviate from the distorted hexagon in Figure \ref{fig:equil_shape}, as edge rounding will take place.

\section{Order-disorder transition}

In his seminal paper Houtappel \cite{Houtappel1950} derived the critical temperature of a triangular Ising spin lattice with anisotropic interactions (labeled $J_{1-3}$). Houtappel found that: $J_1+J_2>0$, $J_2+J_3>0$, and $J_3+J_1>0$, namely without a loss of generality $J_1\geq J_2\geq J_3$, with only one of the terms being potentially allowed to have a negative value. At 0 K any entropy term gives no contribution, thus a boundary free energy is determined by the formation energy of a domain wall segment, and the energy has to be positive. Our analysis of the domain wall free energy directly leads to the conditions as described in the work of Houtappel, especially when we consider the illustrations in Figure \ref{fig:boundary diagram}. In our case we have: $\epsilon_2/2+\epsilon_3/2>0$, and $\epsilon_3/2+\epsilon_1/2>0$ (from segments `23' and `13'). Even though that segment `12' is forbidden at 0 K, it is only because $\epsilon_1/2+\epsilon_2/2$ leads to the most energy expensive segment.

At the order-disorder phase transition temperature $T_c$, the free energy becomes 0.  Thus, the partition sum at the critical temperature is equal to 1. This leads to the expression which gives the order-disorder transition condition. Using the $Z$ derived earlier in Equation (\ref{eq:Z}), the condition reads:

\begin{align} \label{eq:phase transition}
e^{-(\epsilon_1+\epsilon_2)/2kT_c}+ e^{-(\epsilon_2+\epsilon_3)/2kT_c}+ e^{-(\epsilon_3+\epsilon_1)/2kT_c}=1
\end{align}
\smallskip

The expression is symmetric in $\epsilon_1$, $\epsilon_2$, and $\epsilon_3$. This means that we expect an order-disorder phase transition at the same temperature for any of the three high-symmetry directions, just as the free energy becomes zero at the same point in Figure \ref{fig:free energy plot}. In the case of equal interaction energies $\epsilon_1=\epsilon_2=\epsilon_3=\epsilon$, from Equation (\ref{eq:phase transition}) we recover the result of Houtappel \cite{Houtappel1950} and Wannier \cite{Wannier1945}, or that $T_c=\epsilon/k\ln(3)$, proving the validity of our expression. This equation will be discussed in further detail in relation to the work of Houtappel in the next section.

\section{Phase diagram}

In this section we present the phase diagram that can be obtained for the order-disorder phase transition. The phase transition is governed by Equation (\ref{eq:phase transition}) which we obtained with the use of simple statistical mechanics. For the order-disorder point Houtappel obtains a slightly different expression \cite{Houtappel1950}:

\begin{align} \label{eq:phase transition 2}
\sinh(2J_1/kT_c)\sinh(2J_2/kT_c)+ \sinh(2J_2/kT_c)\sinh(2J_3/kT_c)+\\
+\sinh(2J_3/kT_c)\sinh(2J_1/kT_c) =1\nonumber
\end{align}
\smallskip

Note that in the case of Houtappel, $J_{1-3}$ correspond to the spin interaction energies and therefore have a factor of 4 difference compared to our $\epsilon$'s (for details see \cite{Zandvliet2006}). Now, it is obvious that our expression for the order-disorder transition (Equation (\ref{eq:phase transition})) is not analogous to the expression of Houtappel. However, we argue that Equation (\ref{eq:phase transition 2}) needs to be solved with additional conditions in order to satisfy the presented in Houtappel's paper requirements for $J_{1-3}$. For instance, in its current form, Equation (\ref{eq:phase transition 2}) allows for the flipping of the sign of each $J$, without loss of equation validity. This is clearly in contradiction with the condition that maximum one interaction energy can be negative and smaller in magnitude than the rest of the $J$'s. As a result, there are two possible solutions, only one of which will be physically relevant. Thus, when we plot and compare the phase diagrams as obtained by our and Houtappel's models, we need to pose the same conditions that reflect the physical requirements for the interaction energies.

\begin{figure}[ht]
	\centering
	\includegraphics[width=0.6\linewidth]{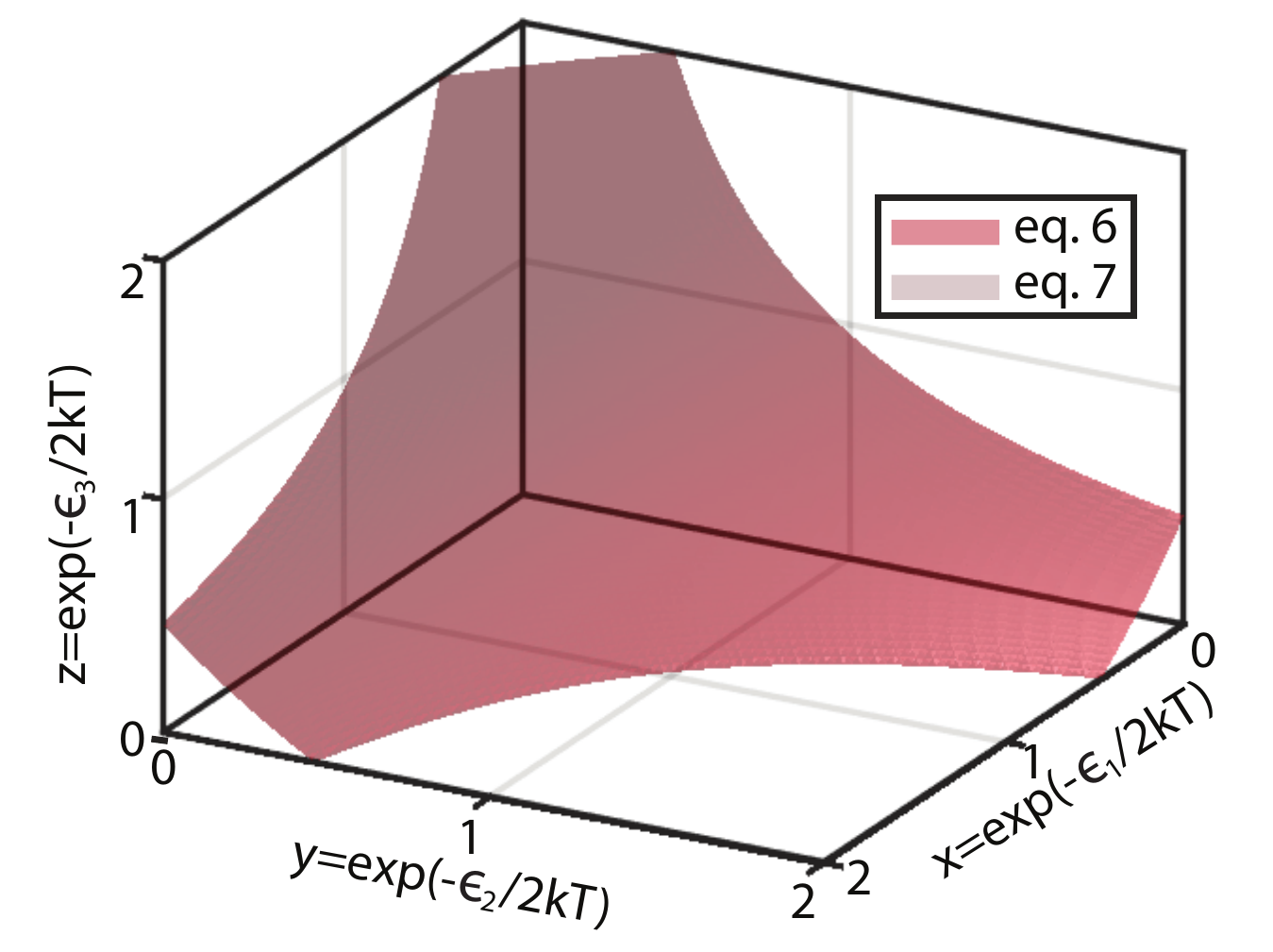}
	\smallskip
	\caption{\small Phase diagram for the domain wall order-disorder transition, presenting an isosurface of the the results in equations (\ref{eq:phase transition}) and (\ref{eq:phase transition 2}) (adapted from Houtappel \cite{Houtappel1950}), plotted in the form of equations (\ref{eq:phase_plot1}) and (\ref{eq:phase_plot2}), respectively. There is a perfect overlap of the isosurfaces (represented here by mixing the colors from the legend).}
	\label{fig:phase diagram}
\end{figure}

Figure \ref{fig:phase diagram} shows the phase diagram, which is an isosurface obtained from equations (\ref{eq:phase transition}) and (\ref{eq:phase transition 2}). We found a perfect overlap of the surface plots, thus an agreement between our solution concerning nearest-neighbor bond strengths and the solution of Houtappel that concerns Ising interactions. On the axes of the phase diagram we have an exponential $e^{-\epsilon_{1-3}/2kT}$, corresponding to $x$, $y$, and $z$ respectively. (In terms of $J$ which is 1/4 times as large as $\epsilon$, for our choice of accounting this becomes $e^{-2J_{1-3}/kT}$). The isosurfaces are then plotted as:

\begin{align}\label{eq:phase_plot1}
xy+yz+zx&=1\ , \\\label{eq:phase_plot2}
1/4(x-1/x)(y-1/y)+ 1/4(y-1/y)(z-1/z)+1/4(z-1/z)(x-1/x)&=1\ ,
\end{align}
\smallskip

\noindent where the first equation corresponds to our solution, while the second corresponds to Houtappel (in a form of a hyperbolic sine expansion in terms of exponents). The conditions that we force are that in all cases $x$, $y$, and $z$ have to remain positive ($x> 0$, $y>0$, $z>0$). Additionally, if one of the interaction energies is allowed to be negative, having the smallest magnitude of the three, this means that the product of any two exponents should remain smaller than 1, thus the next set of conditions is that $xy<1$, $yz<1$, and $zx<1$.

To further discuss the validity of the phase diagram, we set one of the interaction energies to 0. This means we take a cross section from the isosurface, corresponding to either $x$, $y$, or $z$ equal to 1 (a number to the power of 0 equals 1). Then we are able to recover the expected behavior for a 2D square lattice with anisotropic interactions, as derived by Onsager \cite{Onsager1944}. This is not strange. It is already visible from the equation of Houtappel that if we set either of the three $J$'s to 0, then we also obtain the well-known Onsager relation, or $
\sinh(2J_1/kT_c)\sinh(2J_2/kT_c)=1$. This expression is also obtainable from our derivation in Equation (\ref{eq:phase transition}), by setting one of the interaction energies to zero and using the exponential expansion of a hyperbolic sine.

\section{Step meandering}

So far, with the use of the exact expression for the partition sum, it was possible to analyze the behavior of the boundary free energy and to obtain a phase diagram for the order-disorder transition of a triangular lattice. With the use of the partition sum it is also possible to obtain more statistical mechanical quantities, such as the mean squared kink length $\langle n^2 \rangle$ which can be used to obtain the interaction energies in the anisotropic triangular lattice, as demonstrated in this section. Here we treat $\langle n^2 \rangle$ as a dimensionless quantity, however physically it is simply in units of the lattice constant squared, as an individual kink has a length of the lattice constant (see Figure \ref{fig:kink diagram}). The total kink length is simply the number of kinks with respect to the unkinked boundary (at 0 K) multiplied with the individual kink length (the lattice constant). As a dimensionless quantity, the kink length is simply equal to $n$.

To obtain $\langle n^2 \rangle$ the partition sum is used according to the well known statistical mechanics method for obtaining means:

\begin{align}
\langle n^2 \rangle &= \frac{\sum_{n=1}^{\infty} n^2 p_{+n} + \sum_{n=1}^{\infty} (-n)^2 p_{-n}}{Z}
= \frac{\sum_{n=1}^{\infty} n^2 (p_{+n}+p_{-n})}{Z},
\end{align}
\smallskip

\noindent where $p_{+n}$ and $p_{-n}$ correspond to the $n$-th exponent term of the partition sum $Z$; the zeroth order term obviously does not contribute to the sum. Detailed derivation of $\langle n^2 \rangle$ is provided in the Appendix. Here, we give the following final expression for the kinks illustrated as in Figure \ref{fig:kink diagram} (B):

\begin{align} \label{eq:kink squared}
\langle n^2 \rangle = \frac{e^{-\epsilon_1/2kT}e^{-\epsilon_2/2kT}(e^{\epsilon_2/2kT}+e^{-\epsilon_2/2kT})(1+e^{-(\epsilon_1+\epsilon_2)/2kT})}{(e^{-\epsilon_1/2kT}+e^{-\epsilon_2/2kT})(1-e^{-(\epsilon_1+\epsilon_2)/2kT})^2}\ .
\end{align}
\smallskip

Note that the $\epsilon_3$ energy does not participate in the equation, as the kink direction is perpendicular to the $\hat{\epsilon_3}$ segment of the dual lattice (for the choice in Figure \ref{fig:kink diagram} (B), the domain wall is already along the $\hat{\epsilon_3}$ high-symmetry direction). The equation for $\langle n^2 \rangle$ is not symmetric, owing to the asymmetry in the positive and negative kinks (for additional discussion, please see the Appendix).

By simply studying the phase boundary meandering, it is possible to obtain an approximation for the nearest-neighbor interaction energies, given the temperature is known, and from then on, the critical temperature of the order-disorder transition in Equation (\ref{eq:phase transition}). An example is provided in the following section.

$\langle n^2 \rangle$ essentially provides an expectation value for the domain wall width at a given temperature. Way below the critical temperature $T_c$, a domain wall will experience thermal induced meandering. At the transition temperature, although the domain boundary vanishes, one can still use Equation (\ref{eq:kink squared}) to obtain an upper limit for $\langle n^2 \rangle$, and effectively use this limit as an estimate to how close the system is to an order-disorder transition. For instance, in the case of the isotropic triangular lattice ($\epsilon_1= \epsilon_2=\epsilon_3=\epsilon$ and $T_c=\epsilon/k\ln(3)$), one obtains $\langle n^2\rangle_{T_c}=2$.

\smallskip

\section{Application of the model to experimental data}

It is clear that the mean squared kink length is a useful quantity because the expression for $\langle n^2 \rangle$ depends on the interaction energies. Thus, if there is a way to obtain this quantity experimentally, then the interaction energies can be treated as fitting parameters or directly calculated, depending on the system. A way to obtain the mean squared kink length from experimental data is to analyze the boundary between and ordered and a disordered phase. The approach discussed here is in particular suitable for Scanning Tunneling Microscopy (STM) data, for which the resolution is sufficient to study the domain wall between two material phases.

The analysis of a phase boundary or step edge has already been demonstrated with experimental data in a few earlier reports \cite{Kai2013a, Kai2013, Zandvliet1992}. Here we aim at applying our model to the boundary between an ordered and a disordered decanethiol ($C_{10}H_{22}S$) molecular phase on the three-fold symmetric Au(111). Poirier studied in detail the formation of decanethiol self-assembled monolayers on Au(111) and reported the presence of six distinct phases: four ordered phases ($\beta$, $\delta$, $\chi$, and $\phi$) and two disordered phases ($\alpha$ and $\varepsilon$) \cite{Poirier1999}. We choose to analyze the boundary between the ordered $\beta$ phase and the disordered $\varepsilon$ phase. The domain wall separating these two must be experiencing, or be very close, to an order-disorder phase transition as was reported that the $\beta$ phase can remain stable only up to 350 K \cite{Poirier2001}.

We proceed to our model of the molecular phase ordering. Note that we discuss it within the framework of the triangular lattice, and therefore we define three effective interaction energies, as the symmetry of the $\beta$ phase suggests. In the current context, we do not distinguish the different physical aspects of all the possible interactions between the molecules and between the molecules and the underlying substrate. We provide the current analysis as an illustration of how our model of the anisotropic triangular lattice can be applied to experimental data.

A model of the $\beta$ phase and its boundary is provided in Figure \ref{fig:decanethiol} (A). Again we use the same dual lattice system to mimic the Au(111) substrate: the Au atoms locations are in the center of each honeycomb cell. For the molecular phase we have selected a unit cell that contains two molecules. As a result, the lattice is a triangular lattice with one of its dimensions longer than the other two. This is a consequence of the strong anisotropy in the interaction between molecules. The three interaction energies that we define are shown with arrows: two interactions of the same strength between neighboring molecules ($\epsilon_1/2=\epsilon_2/2$ shown in blue, at the tail region of the molecules), and a third, strongest interaction energy along neighboring molecules ($\epsilon_3/2$ shown with a red arrow). We selected two interaction energies of the same strength as according to our diagram, only one of the three nearest-neighbor distances is differing from the other two. In the inset of (A) we have also shown the molecular phase ordering idea if the anisotropy in the molecular lattice dimensions is neglected (again the honeycomb dual lattice is shown). In Figure \ref{fig:decanethiol} (B), we present the respective STM experimental data. A full correspondence is made to the model in (A): at each molecular row the boundary is marked with a blue circle, so at each molecular row the domain wall location is known. We measure the domain wall location with respect to a reference line (in blue) and after that we are able to determine the amount of molecules present from the reference to each boundary point. With the blue dashed line in Figure \ref{fig:decanethiol} (A) the unkinked domain wall is outlined (based on the expectation that at 0 K, only two interaction energies contribute to the free energy, see the discussion on the free energy above). With respect to that boundary we are able to account for the amount of molecules extra (or less) at each molecular row.

There is a direct correlation between the amount of extra/less molecules with respect to the unkinked boundary and the amount of kinks (for a detailed description of the mean squared kink length expression and the counting of kinks, please see the Appendix). At each molecule location, we determine the kink length with respect to the previous position. For instance, in Figure \ref{fig:decanethiol} (A), beginning from the location marked with a `*', we count 0, +2, 0 molecules. This translates to 0, +2, -2 kinks. This counting system should be propagated along the whole boundary. Finally, from the kink length distribution we are able to calculate $\langle n^2 \rangle$, by simply using the statistical method of obtaining a mean. The counts from analyzing multiple boundaries as the one illustrated in Figure \ref{fig:decanethiol} (B) are shown in Table \ref{tab:nsquare}.

We obtained for the mean squared kink length $\langle n^2 \rangle$ a value of 1.96. We can now equate this value to Equation (\ref{eq:kink squared}). Note that, based on the explanation in the previous section and the interaction energies labeled in Figure \ref{fig:decanethiol} (A), Equation (\ref{eq:kink squared}) is reduced to the following (because $\epsilon_1/2=\epsilon_2/2$):

\begin{align}\label{eq:nsqbeta}
\langle n^2 \rangle &= \frac{(1+e^{-\epsilon_1/kT})^2}{2(1-e^{-\epsilon_1/kT})^2}\ . 
\end{align}
\smallskip

\begin{figure}[ht]
	\centering
	\includegraphics[width=0.6\linewidth]{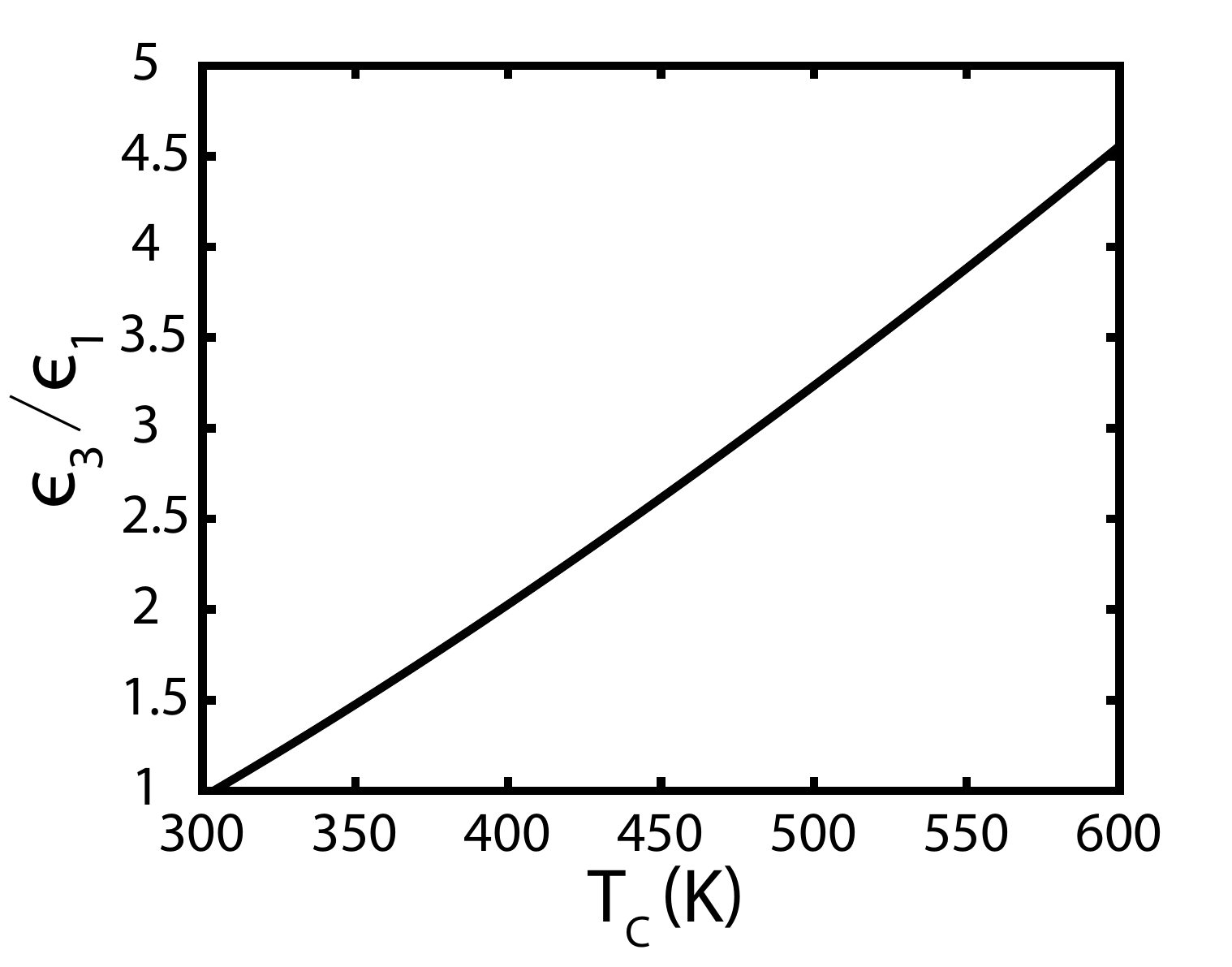}
	\caption{\small Calculated order-disorder temperature $T_c$ (x axis) by using Equation (\ref{eq:phase transition}) and Equation (\ref{eq:nsqbeta}),  for a range of ratios of $\epsilon_3/\epsilon_1$ (on the y axis). $\epsilon_2=\epsilon_1$, $\langle n^2 \rangle=1.96$.}
	\label{fig:tc_gamas}
\end{figure}

Having obtained $\epsilon_1=\epsilon_2=29$ meV (for $T=300$ K), we need to evaluate the third interaction energy in order to estimate the order-disorder transition temperature. However, for that we need to analyze a boundary of the type marked with a pink dashed line in Figure \ref{fig:decanethiol} (A). Such boundaries were always too short in the experimental data and not suitable to perform proper analysis. When looking at the equilibrium shape in Figure \ref{fig:equil_shape} (knowing that two interaction energies are equal) we would expect two boundaries which are a lot longer at room temperature, and one very short boundary. Another explanation emerges from consulting with Figure \ref{fig:free energy plot}. We notice that close to the order-disorder temperature the free energies of all three domain walls in a triangular lattice become very close to each other. Therefore, it would not be possible to distinguish nicely any longer the three domain walls of a triangular island, or it becomes unclear which segments of the decanethiol island domain walls belong to the boundary type that we analyzed above, and which are kinked segments of the other, shorter boundary. Of course, this would strongly depend on the difference between the interaction energies. That is why, in Figure \ref{fig:tc_gamas} we decided to plot the transition temperature $T_c$ based on a range of possible ratios of $\epsilon_3/\epsilon_1$. According to the report of Poirier \cite{Poirier2001}, and the fact that the $\beta$ phase is stable up to 350 K, it seems that the third interaction energy should be only about 1.5 times larger than the other two.

One can argue if the $\epsilon_3$ energy can be only 50 \% larger than the other two interaction energies, considering the very strong anisotropy in the molecular ordering. To that end, we suggest that our model can be improved via the inclusion of next-nearest-neighbor interactions. Additionally, here we have not considered that for the transition of the $\beta$ phase to a disordered state, also the interaction between molecules within the same unit cell must be overcomed (here we take two molecules as one unit). Nevertheless, these discrepancies only teach us how to improve our model, through them we could learn even more about the decanethiol $\beta$ phase: although approached as anisotropic triangular system here, the seemingly simple ordering of that phase is far from trivial. The molecules assume ordering in long molecular rows which are also three-fold symmetric with respect to the underlying substrate, thus, not only the next-nearest-neighbor interactions, but also the monolayer--substrate interaction is crucial. We are confident that only analysis concepts as the one presented here can tackle the still unanswered questions regarding the self-assembly of such molecular layers in the future. The complexity of such systems is additionally increased due to the effect of neighboring phases, and the generally high molecular mobility and monolayer dynamics. These effects, of course, fall beyond the analysis capabilities of a lattice model.

\section{Discussion}

The partition sums for the domain walls of a triangular lattice with anisotropic nearest-neighbor interactions were derived exactly. Using these expressions, we showed that the triangular lattice undergoes an order-disorder transition at the same critical temperature, governed by Equation (\ref{eq:phase transition}), given a proper choice for the interaction energies $\epsilon_1+\epsilon_2>0$, $\epsilon_2+\epsilon_3>0$, $\epsilon_3+\epsilon_1>0$, equivalent to the conditions posed by Houtappel in his work on the phase transition of the 2D triangular Ising lattice \cite{Houtappel1950}. It should be pointed out that our method, which relies on the solid-on-solid model, does not, in general, provide exact solutions for the domain boundary free energy and the order-disorder temperature. However, in the case of the two-dimensional square and triangular lattices with only nearest-neighbor interactions (isotropic or anisotropic) our results are exactly the same as the results obtained by Onsager \cite{Onsager1944} and Houtappel \cite{Houtappel1950}, respectively. A phase diagram for the order-disorder transition is presented and also features an overlap with the findings of Houtappel when proper conditions are set to both equations (\ref{eq:phase transition}) and (\ref{eq:phase transition 2}). When one of the interaction energies is set to zero, the Onsager relation for the 2D Ising lattice is recovered. The domain wall free energy was studied for the three high-symmetry directions of the triangular lattice and the results show that in the case of anisotropic nearest-neighbor interactions, there will be preferred domain walls. A method for studying the thermal induced domain wall meandering was also developed by using a statistical mechanics approach to deriving the mean squared kink length for the boundary that separates a material from a neighboring phase. That way, the theoretical model presented in this paper can be applied to study experimental data involving materials with triangular symmetry in two dimensions. Our boundary meandering approach is able to produce approximations for the nearest-neighbor interaction energies when the temperature is known, or can be used to determine the critical temperature of the order-disorder transition when information is available about the interaction energies. In this paper we applied this method to study the order-disorder transition for a phase of decanethiol molecules. For two high-symmetry directions we could obtain the mean squared kink length. This provided us with the opportunity to obtain an estimate for the order-disorder transition temperature versus the ratio of the third interaction energy with respect to the other two. The currently discussed framework is, of course, not limited to the investigation of STM data only, given there is sufficient resolution to analyze a boundary. As an outlook, to obtain even better estimates of the transition temperature and to analyze in greater detail anisotropic triangular systems, it would be required to add next-nearest-neighbor interactions to the model. To analyze a particular molecular system in every aspect, our simple solid-on-solid model has, of course, its limitations.

\newpage
\setcounter{equation}{0}
\renewcommand{\theequation}{A.\arabic{equation}}
\section{Appendix}

Here we present the derivations of several quantities addressed in the main text and provide further discussion when applicable. 

\bigskip
\subsection{Derivation of the partition sum}

The partition sum can be derived by accounting for the kink formation energies in Figure \ref{fig:kink diagram} from the main text and has the form:

\begin{align}
Z= p_{0}+ \sum_{n=1}^{\infty} (p_{-n} + p_{+n})\ ,
\end{align}
\smallskip

\noindent where the $p$-terms have the form $e^{-E/kT}$, with the $E$ corresponding to the kink formation energy (or sum of segments of the honeycomb dual lattice in Figure \ref{fig:kink diagram} from the main text). Let's look at the zeroth order kink and several positive and negative kinks:

\begin{align}\label{eq:p_exponents}
p_0&= \exp(-(\epsilon_2/2+\epsilon_3/2)/kT)= \exp(-(\epsilon_2+\epsilon_3)/2kT)\ ,\\
p_1&= \exp(-(\epsilon_1/2+\epsilon_3/2)/kT)= \exp(-(\epsilon_1+\epsilon_3)/2kT)\ ,\\
p_2&= \exp(-(2\epsilon_1/2+\epsilon_2/2+ \epsilon_3/2)/kT)= \exp(-(2\epsilon_1+\epsilon_2+\epsilon_3)/2kT)\ ,\\
p_{-1}&= \exp(-(\epsilon_1/2+2\epsilon_2/2+ \epsilon_3/2)/kT)= \exp(-(\epsilon_1+2\epsilon_2+\epsilon_3)/2kT)\ ,\\
p_{-2}&= \exp(-(2\epsilon_1/2+3\epsilon_2/2+ \epsilon_3/2)/kT)= \exp(-(2\epsilon_1+3\epsilon_2+\epsilon_3)/2kT)\ .
\end{align}
\smallskip

It is easily noticed that the positive kinks then contribute with a sum of the form $\sum_{n=0}^{\infty}  e^{-(n\epsilon_2+(n+1)\epsilon_1)/2kT} e^{-\epsilon_3/2kT}$. If we combine the zeroth term with the negative kinks contributions, then we obtain a sum of the form $\sum_{n=0}^{\infty} e^{-(n\epsilon_1+(n+1)\epsilon_2)/2kT} e^{-\epsilon_3/2kT}$.
The partition sum obtained by accounting for all the positive and the negative kinks then reads:

\begin{align}
Z= e^{-\epsilon_3/2kT}\Big[ \sum_{n=0}^{\infty} e^{-(n\epsilon_1+(n+1)\epsilon_2)/2kT}+ e^{-(n\epsilon_2+(n+1)\epsilon_1)/2kT}\Big]\ .
\end{align}
\smallskip

Now we simplify:

\begin{align}
a= e^{-\epsilon_1/2kT}\ ,\\
b= e^{-\epsilon_2/2kT}\ ,\\
c= e^{-\epsilon_3/2kT}\ ,\\
Z&= c\Big[ \sum_{n=0}^{\infty} a^nb^{n+1} + \sum_{n=0}^{\infty} a^{n+1}b^{n} \Big]\\\nonumber
&= c\Big[ b\sum_{n=0}^{\infty} a^nb^{n} + a\sum_{n=0}^{\infty} a^{n}b^{n} \Big]\\\nonumber
&= c\Big[ (a+b)\sum_{n=0}^{\infty} a^nb^{n} \Big]\\\nonumber
&= e^{-\epsilon_3/2kT}(a+b)\Big[ \sum_{n=0}^{\infty} a^{n}b^{n} \Big]\\\nonumber
&= e^{-\epsilon_3/2kT}(e^{-\epsilon_1/2kT}+e^{-\epsilon_2/2kT}) \sum_{n=0}^{\infty} \Big(e^{-(\epsilon_1+\epsilon_2)/2kT}\Big)^{n} \\
&= \frac{e^{-\epsilon_3/2kT}(e^{-\epsilon_1/2kT}+e^{-\epsilon_2/2kT})}{ 1-e^{-(\epsilon_1+\epsilon_2)/2kT} }= \frac{c(a+b)}{1-ab}, 
\end{align}
\smallskip

\noindent where in the last row we have used that $\sum_0^{\infty} x^n= 1/(1-x)$. Note that this partition sum holds only for the diagram in Figure \ref{fig:kink diagram} from the main text. For the two other high-symmetry directions, the indices of the interaction energies $\epsilon$ need to be permuted. 

The expression for the partition function allows us to calculate additional statistical mechanics quantities. These include, for example, the mean squared kink length (derivation is shown later in this appendix), the expectation value for the kink-kink separation distance $\langle S \rangle$, and the probability distribution of the kink-kink separation $P(S)$. $\langle S \rangle$ and $P(S)$, for instance, read:

\begin{align}
\langle S \rangle &= \sum_{s=1}^{\infty} sq_0^S(1-q_0)^2\ ,\\
P(S) &= q_0^S(1-q_0)^2\ ,
\end{align}
\smallskip

where $q_0$ is equal to $p_0/Z$. $p_0$ was defined in Equation (\ref{eq:p_exponents}). For a given high-symmetry direction the energy indices in the exponent $p_0$ will have to be properly permuted.

\bigskip
\subsection{Derivation of the free energy}

To obtain the free energies for each of the three high-symmetry directions as in the main text, we make the choice that $\epsilon_1>\epsilon_2>\epsilon_3$. We start with the well-known expression from statistical mechanics:

\begin{align}
F&=-kT\ln Z\\
&=-kT\ln\frac{e^{-\epsilon_3/2kT}(e^{-\epsilon_1/2kT}+e^{-\epsilon_2/2kT})}{ 1-e^{-(\epsilon_1+\epsilon_2)/2kT} }\ .
\end{align}
\smallskip

Then we can rewrite the logarithm as:

\begin{align}
F=-kT\ln{e^{-\epsilon_3/2kT}(e^{-\epsilon_1/2kT}+e^{-\epsilon_2/2kT})}+kT\ln{ (1-e^{-(\epsilon_1+\epsilon_2)/2kT}) }\ .
\end{align}
\smallskip

Finally we also simplify the first logarithm term:

\begin{align}
F&=-kT\ln{e^{-\epsilon_3/2kT}}-kT\ln{\Big(e^{-\epsilon_1/2kT}+e^{-\epsilon_2/2kT}\Big)}+kT\ln{ \Big(1-e^{-(\epsilon_1+\epsilon_2)/2kT}\Big) }\\\nonumber
&=-kT\ln{e^{-\epsilon_3/2kT}}-kT\ln e^{-\epsilon_2/2kT}{\Big(1+e^{-\epsilon_1/2kT} e^{\epsilon_2/2kT}\Big)}+kT\ln{ \Big(1-e^{-(\epsilon_1+\epsilon_2)/2kT}\Big) }\\\nonumber
&=\frac{\epsilon_3}{2} + \frac{\epsilon_2}{2} -kT\ln{\Big(1 +e^{-\epsilon_1/2kT} e^{\epsilon_2/2kT}\Big)}+kT\ln{ \Big(1-e^{-(\epsilon_1+\epsilon_2)/2kT}\Big) }\\
&= \frac{\epsilon_3+\epsilon_2}{2} -kT\ln{\Big(1 +e^{(\epsilon_2-\epsilon_1)/2kT}\Big)}+kT\ln{ \Big(1-e^{-(\epsilon_1+\epsilon_2)/2kT}\Big) }\ .
\end{align}
\smallskip

For the other two high-symmetry directions we only permute the indices, but also keep in mind the choice we made at the beginning. At the first logarithm, the exponent is raised to a negative power and does not diverge at $T=0$.

\bigskip
\subsection{Equilibrium shape at 0 K}

The analysis we have provided so far on the free energy of course applies to a freely meandering boundary. We now want to derive the equilibrium island shape of the triangular lattice with anisotropic nearest-neighbor interactions. We restrict ourselves to $T= 0$ K. At higher temperature, due to the known expression for the free energy $F= U- TS$, the second (entropy) term gives contribution and then the calculation becomes more complex (not shown here), as then one needs to also obtain the domain wall free energy along an arbitrary direction (not only the principle directions with which we deal in the current report).

To construct the equilibrium shape at 0 K, we need to obtain a polar plot of the energy $\gamma(\theta)$ and then perform a Wulff construction. The polar plot and the Wulff construction were presented in Figure \ref{fig:equil_shape} from the main text. Here we focus on how the polar plot was obtained.

\begin{figure}[ht]
	\centering
	\includegraphics[width=0.7\linewidth]{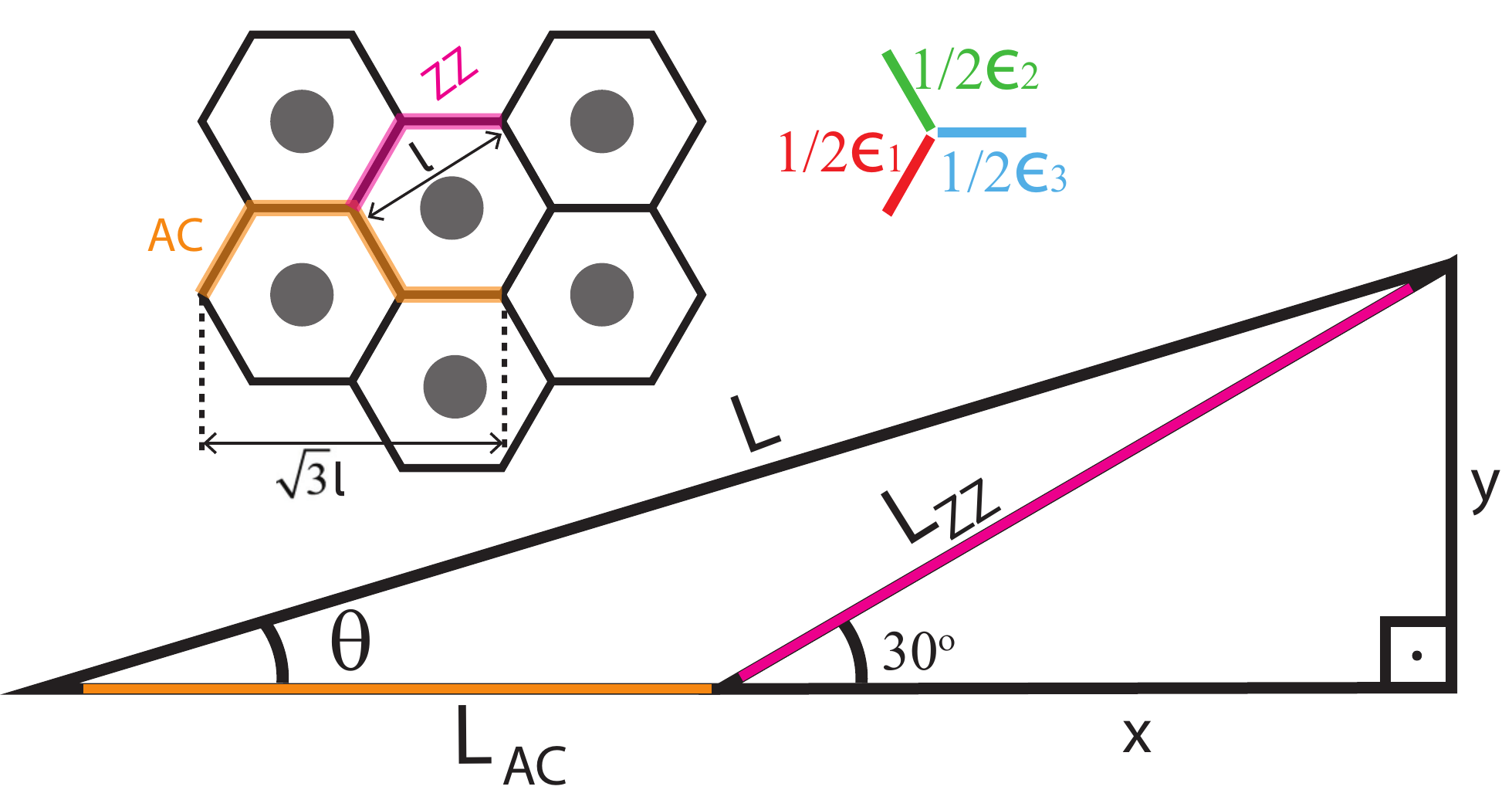}
	\caption{\small Illustration supporting the equilibrium shape calculation. The triangular lattice (gray circles) and the dual lattice (honeycomb grid) are shown at the top left. The unit length is denoted by $l$. In orange an armchair (AC) type of segment from the dual lattice is shown, with length of $\sqrt{3}l$. With magenta a zig-zag (ZZ) type of segment from the dual lattice is shown, with a length $l$. A legend of the energies corresponding to the individual honeycomb segments is shown at the top right. Below, a triangle is constructed, showing the contribution of AC and ZZ types of segments in a total segment L. The triangle is used to calculate the energy $\gamma(\theta)$, as explained in the Appendix.}
	\label{fig:ac_zz_0k}
\end{figure}

In Figure \ref{fig:ac_zz_0k} we present an illustration of how the energy $\gamma(\theta)$ can be obtained. We distinguish between the general polar angle $\phi$ which runs between $0^{\circ}$ and $360^{\circ}$ (between 0 and $2\pi$) and $\theta$ which is always between $0^{\circ}$ and $30^{\circ}$ (between 0 and $\pi/6$) because once the energy $\gamma(\theta)$ is obtained, $\gamma(\phi)$ follows with a few adjustments that we will explain below. We again make extensive use of the honeycomb dual lattice for simplicity. When we look along a $\theta=0^{\circ}$ angle in our diagram, a unit segment of the dual honeycomb lattice corresponds to an armchair segment (AC) marked with orange. When one looks along a $\theta=30^{\circ}$ ($\pi/6$) angle, a unit segment corresponds to a zig-zag (ZZ) segment of the dual honeycomb lattice, marked in magenta. The energy corresponds to $\gamma_{\text{AC}}^{0}= (\epsilon_1+\epsilon_2+2\epsilon_3)/(2\sqrt{3})$ per unit length for the AC segment, while it is $\gamma_{\text{ZZ}}^{\pi/6}=(\epsilon_1+\epsilon_3)/2$ per unit length for the ZZ segment. Any arbitrary segment of length L can be broken down in contributions of lengths along the AC (L$_{\text{AC}}$) and the ZZ (L$_{\text{ZZ}}$) directions by using the triangle:

\begin{align}
y&= \text{L}_{\text{ZZ}} \sin(30^{\circ})= \frac{1}{2} \text{L}_{\text{ZZ}}= \text{L}\sin(\theta)\ \implies \ \frac{\text{L}_{\text{ZZ}}}{\text{L}}=2\sin{\theta}\ , \\\nonumber 
\cos(\theta)&= \frac{\text{L}_{\text{AC}} + x}{\text{L}}= \frac{\text{L}_{\text{AC}}+\text{L}_{\text{ZZ}} \cos(30^{\circ})}{\text{L}}= \frac{\text{L}_{\text{AC}}+ \frac{\sqrt{3}}{2}\text{L}_{\text{ZZ}}}{\text{L}}\ ,\\
&\implies \frac{\text{L}_{\text{AC}}}{\text{L}}= \cos(\theta)-\sqrt{3}\sin(\theta)\ . 
\end{align}
\smallskip

Now we are ready to calculate the free energy $\gamma(\theta)$. $\gamma(\phi)$ between $0^{\circ}$ and $30^{\circ}$ (from 0 to $\pi/6$) directly follows. This is also the portion of the polar plot between 0 and $\pi/6$ in Figure \ref{fig:equil_shape} from the main text: 

\begin{align}
\gamma(\theta)= \gamma(\phi)_{0\rightarrow\pi/6} &= \frac{\text{L}_{\text{AC}}}{\text{L}}\gamma_{\text{AC}}^{0} + \frac{\text{L}_{\text{ZZ}}}{\text{L}}\gamma_{\text{ZZ}}^{\pi/6}\\\nonumber
&= [\cos(\theta)-\sqrt{3}\sin(\theta)] \gamma_{\text{AC}}^{0} + 2\sin{\theta}\gamma_{\text{ZZ}}^{\pi/6}
\end{align}
\smallskip

It is possible to obtain the free energy $\gamma$ for the general angle $\phi$ if we use the equations we derived so far. Consulting with the triangle in Figure \ref{fig:ac_zz_0k}, it is clear that the L$_{\text{AC}}$ and L$_{\text{ZZ}}$ segments are interchangeable in the analysis. In fact, each $30^{\circ}$ ($\pi/6$) of the general polar angle $\phi$ they must be swapped. This leads to the fact that the formula in our last equation can be generalized and applied in $30^{\circ}$ ($\pi/6$) portions in order to obtain the full polar plot. The only thing that has to be carefully changed each time is the energies $\gamma_{\text{AC,ZZ}}$ because after a rotation of $\pi/6$ due to the anisotropy of the lattice, the energy per unit AC/ZZ segment changes. The unique $\gamma_{\text{AC,ZZ}}$ are listed below, while the formulas for obtaining the full polar plot of the free energy $\gamma$ at 0 K are given for each portion of the polar angle $\phi$ in Table \ref{tab:polar_plot}.

\begin{align}
\gamma_{\text{AC}}^{0^{\circ}}&= \gamma_{\text{AC}}^{180^{\circ}}= \gamma_{\text{AC}}^{0}= \frac{\epsilon_1+\epsilon_2+2\epsilon_3}{2\sqrt{3}}\ , \\
\gamma_{\text{AC}}^{60^{\circ}}&= \gamma_{\text{AC}}^{240^{\circ}}=\gamma_{\text{AC}}^{\pi/3}= \frac{2\epsilon_1+\epsilon_2+\epsilon_3}{2\sqrt{3}}\ , \\
\gamma_{\text{AC}}^{-60^{\circ}}&= \gamma_{\text{AC}}^{120^{\circ}}= \gamma_{\text{AC}}^{5\pi/3}= \frac{\epsilon_1+2\epsilon_2+\epsilon_3}{2\sqrt{3}}\ ,\\
\gamma_{\text{ZZ}}^{30^{\circ}}&= \gamma_{\text{ZZ}}^{210^{\circ}}= \gamma_{\text{ZZ}}^{\pi/6}= \frac{\epsilon_1+\epsilon_3}{2}\ , \\
\gamma_{\text{ZZ}}^{90^{\circ}}&= \gamma_{\text{ZZ}}^{-90^{\circ}}=\gamma_{\text{ZZ}}^{\pi/2}= \frac{\epsilon_1+\epsilon_2}{2}\ , \\
\gamma_{\text{ZZ}}^{-30^{\circ}}&= \gamma_{\text{ZZ}}^{150^{\circ}}= \gamma_{\text{ZZ}}^{11\pi/6}= \frac{\epsilon_2+\epsilon_3}{2}\ ,
\end{align}
\smallskip

Finally, we stress upon the fact again that the AC/ZZ segments correspond to portions of the honeycomb dual lattice and were used for simplicity. There segments do not have physical meaning when it comes to the actual domain boundary.

\bigskip
\subsection{Derivation of the mean squared kink length}

Analogous to the derivation of the partition sum, to obtain $\langle n^2 \rangle$, the aim is to obtain a sum of the type: 

\begin{align}
\langle n^2 \rangle = \frac{\sum_{n=1}^{\infty} n^2 (p_{+n}+p_{-n})}{Z}= \frac{P_{+n}+P_{-n}}{Z}\ .
\end{align}
\smallskip

Luckily we already obtained the partition sum. The derivation for $\langle n^2 \rangle$ is analogous, only now $n^2$ need to be multiplied with the exponent terms. Let's look at several positive and negative kink contributions; note that the zeroth order kink (or the no-kink) gives no contribution here:

\begin{align}
P_{0}&=0^2 p_0= 0\ ,\\
P_{1}&=1^2p_1= \exp(-(\epsilon_1+\epsilon_3)/2kT)\ ,\\
P_{2}&=2^2p_2= 2^2 \exp(-(2\epsilon_1+\epsilon_2+\epsilon_3)/2kT)\ ,\\
P_{-1}&=(-1)^2p_{-1}= \exp(-(\epsilon_1+2\epsilon_2+\epsilon_3)/2kT)\ ,\\
P_{-2}&=(-2)^2p_{-2}= 2^2\exp(-(2\epsilon_1+3\epsilon_2+\epsilon_3)/2kT)\ .
\end{align}
\smallskip

From the pattern visible in the $P$-terms it becomes clear that:

\begin{align}
P_{+n}&=  \sum_{n=1}^{\infty} n^2 e^{-(n\epsilon_1+(n-1)\epsilon_2)/2kT}e^{-\epsilon_3/2kT}\ ,\\
P_{-n}&=  \sum_{n=1}^{\infty} n^2 e^{-(n\epsilon_1+(n+1)\epsilon_2)/2kT}e^{-\epsilon_3/2kT}\ .
\end{align}
\smallskip

First we simplify the expression for $P_{+n}+P_{-n}$.

\begin{align}
P_{+n} + P_{-n} &=  \sum_{n=1}^{\infty} n^2 e^{-(n\epsilon_1+(n-1)\epsilon_2)/2kT}e^{-\epsilon_3/2kT}
+  \sum_{n=1}^{\infty} n^2 e^{-(n\epsilon_1+(n+1)\epsilon_2)/2kT}e^{-\epsilon_3/2kT}\\
&= e^{-\epsilon_3/2kT}\Big[ \sum_{n=1}^{\infty} n^2 e^{-(n\epsilon_1+(n-1)\epsilon_2)/2kT}
+  \sum_{n=1}^{\infty} n^2 e^{-(n\epsilon_1+(n+1)\epsilon_2)/2kT}\ \Big]\ .
\end{align}
\smallskip

Again making use of substitution by letters we obtain:

\begin{align}
a= e^{-\epsilon_1/2kT}\ ,\\
b= e^{-\epsilon_2/2kT}\ ,\\
c= e^{-\epsilon_3/2kT}\ ,
\end{align}
\begin{align}
P_{+n} + P_{-n} &= c\Big[ \sum_{n=1}^{\infty} n^2 a^n b^{(n-1)}
+  \sum_{n=1}^{\infty} n^2 a^n b^{(n+1)} \Big]\\\nonumber
&= c\Big[ (1/b+b)\sum_{n=1}^{\infty} n^2 a^n b^{n} \Big]\\
&= c\Big[ (1/b+b)\sum_{n=1}^{\infty} n^2 (ab)^{n} \Big]
= c (1/b+b)\frac{ab(1+ab)}{(1-ab)^3},
\end{align}
\smallskip

\noindent where in the last row it is used that $\sum_{n=1}^{\infty} n^2 x^n= {x(1+x)}/{(1-x)^3}$. Now, to obtain $\langle n^2 \rangle$, we only need to divide by $Z= c(a+b)/(1-ab)$:

\begin{align}
\langle n^2 \rangle&= \frac{P_{+n}+P_{-n}}{Z}\\\nonumber
&= \frac{c (1/b+b)\frac{ab(1+ab)}{(1-ab)^3}}{\frac{c(a+b)}{1-ab}}\\\nonumber
&= \frac{ab(1/b+b)(1+ab)}{(a+b)(1-ab)^2}\\
&= \frac{e^{-\epsilon_1/2kT}e^{-\epsilon_2/2kT}(e^{\epsilon_2/2kT}+e^{-\epsilon_2/2kT})(1+e^{-(\epsilon_1+\epsilon_2)/2kT})}{(e^{-\epsilon_1/2kT}+e^{-\epsilon_2/2kT})(1-e^{-(\epsilon_1+\epsilon_2)/2kT})^2}\ .
\end{align}
\smallskip

\begin{figure}[ht]
	\centering
		\includegraphics[width=0.6\linewidth]{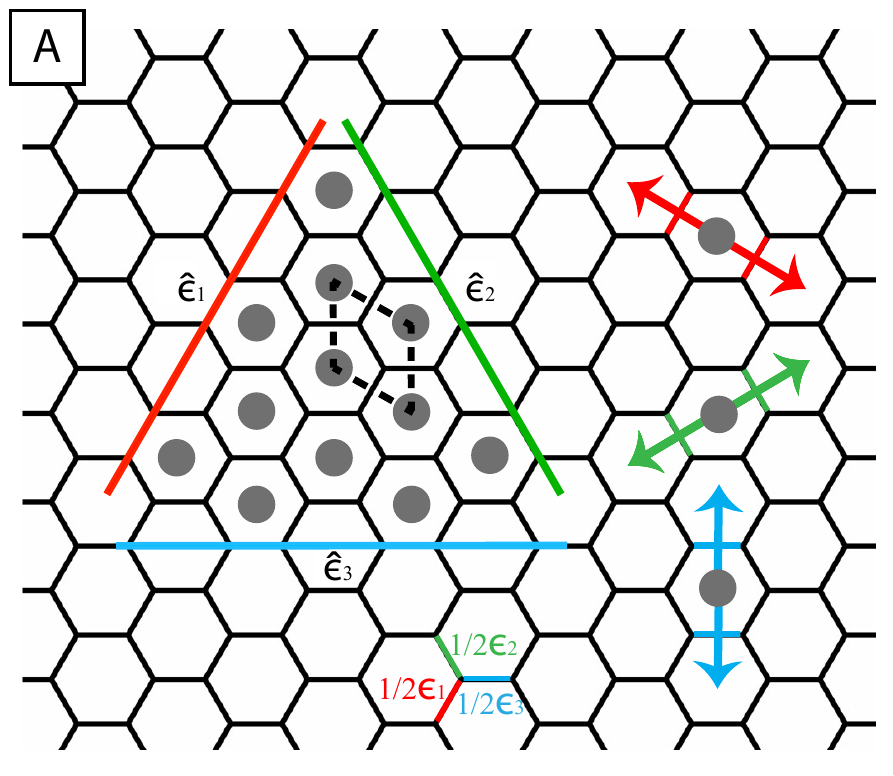}
		\includegraphics[width=0.45\linewidth]{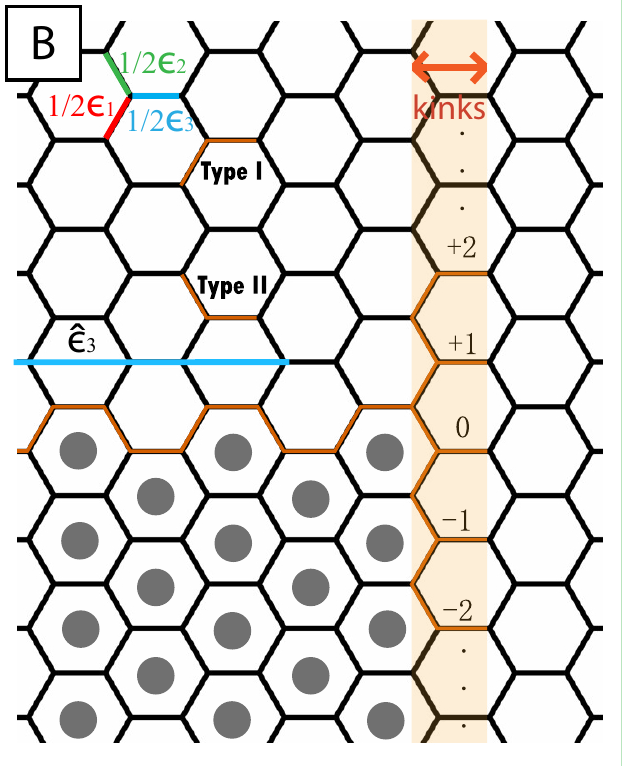}
	\caption{\small Illustration of the dual lattice system. (A) The honeycomb lattice represents the boundaries between atoms, while the gray circles represent the lattice positions of the triangular lattice (black dashed lines show a possible primitive unit cell choice). Color coded are shown the three high-symmetry directions of the triangular lattice: $\hat{\epsilon_1}$ is drawn in red at 60$^{\circ}$ w.r.t. the horizontal direction, $\hat{\epsilon_2}$ is drawn in green at 120$^{\circ}$ w.r.t. the horizontal direction, $\hat{\epsilon_3}$ is drawn horizontally in blue. The high-symmetry directions run perpendicular to the bonds with strength $\epsilon_1$, $\epsilon_2$, and $\epsilon_3$ between atoms, shown as arrows to the right, a single broken bond corresponds to half of the respective bond strength. (B) Illustration of the triangular lattice boundary interrupted by kinks. At the boundary, each kink segment corresponds to a broken bond along one of the three high-symmetry directions, and has a load of either $\epsilon_1/2$, $\epsilon_2/2$, or $\epsilon_3/2$. The uninterrupted boundary is made of repetition of Type I and Type II segments. Each positive kink ends with a Type I segment. Each negative kink ends with a Type II segment. The kink width is constant (shown with an orange arrow). The kink length (height) corresponds to the number of the kink times the corresponding lattice constant which is equal to the height of the honeycomb cell (vertical distance between gray circles).}
	\label{fig:kink diagram}
\end{figure}

\begin{figure}[ht]
	\centering
	\includegraphics[width=0.6\linewidth]{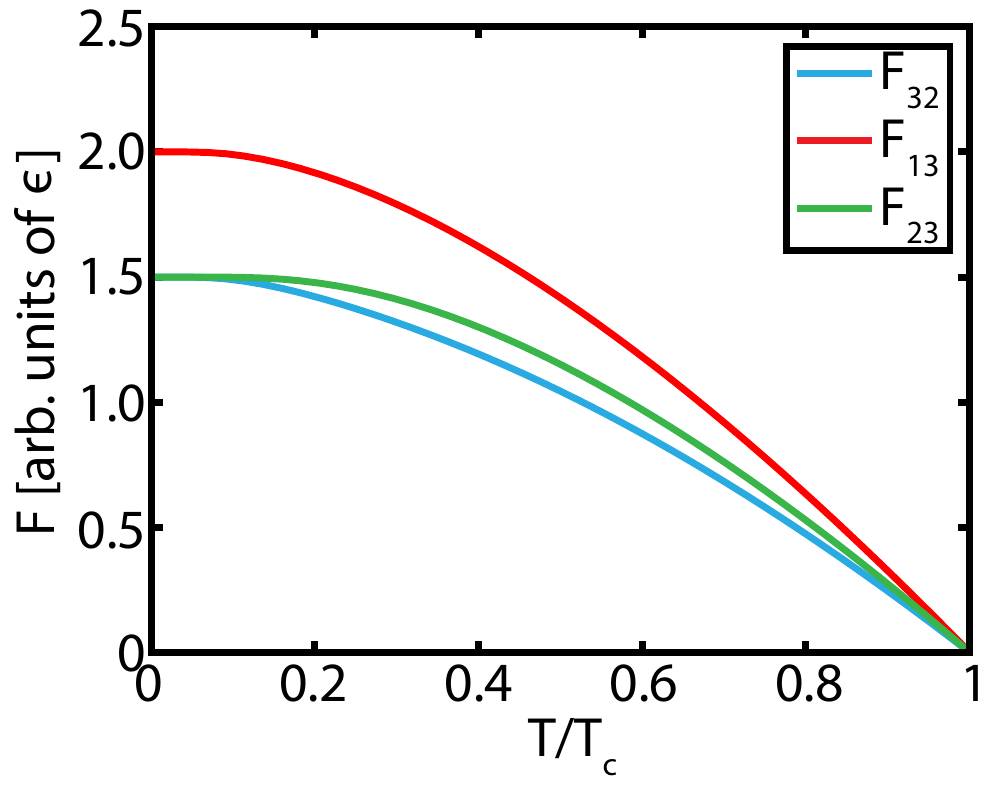}
	\smallskip
	\caption{\small Free energy plot. $F_{\hat{\epsilon_3}}= F_{32}$ (bottom, blue line), $F_{\hat{\epsilon_1}}= F_{13}$ (top, red line), and $F_{\hat{\epsilon_2}}= F_{23}$ (middle, green line), the color code and the high-symmetry directions correspond to Figure \ref{fig:kink diagram}. As the free energy is derived by using $F=-kT\ln Z$, the curves correspond to the free energy per unit length of the domain wall. The choice made is $\epsilon_1>\epsilon_2>\epsilon_3$, or $\epsilon_1= 3\epsilon$, $\epsilon_2=2\epsilon$, $\epsilon_3=\epsilon$ for an arbitrary $\epsilon$.}
	\label{fig:free energy plot}
\end{figure}

The expression is, as expected, not symmetric in $\epsilon_1$ and $\epsilon_2$, owing to the asymmetry of positive and negative kinks. This becomes clear from Figure \ref{fig:kink diagram} (B) from the main text: all positive kinks end with a type I segment, all negative kinks end with a type II segment, but because there is a 0 kink, the positive and negative kinks are not mirrored versions of themselves (with respect to an imaginary horizontal line). This asymmetry is also the reason why it seems to matter if a boundary is analyzed in one direction or the opposite: it would seem that an unkinked boundary is made out of 0 kinks in one direction, but out of +1 kinks in the other direction. However, there should be no outcome degeneracy, irrespective how the high-symmetry direction is selected. Here, we suggest the procedure to overcome this discrepancy. Note that there is no symmetry problem when obtaining the partition sum, because then the contribution of the 0 kink is not excluded!

\begin{figure}[ht]
	\centering
	\includegraphics[width=0.65\linewidth]{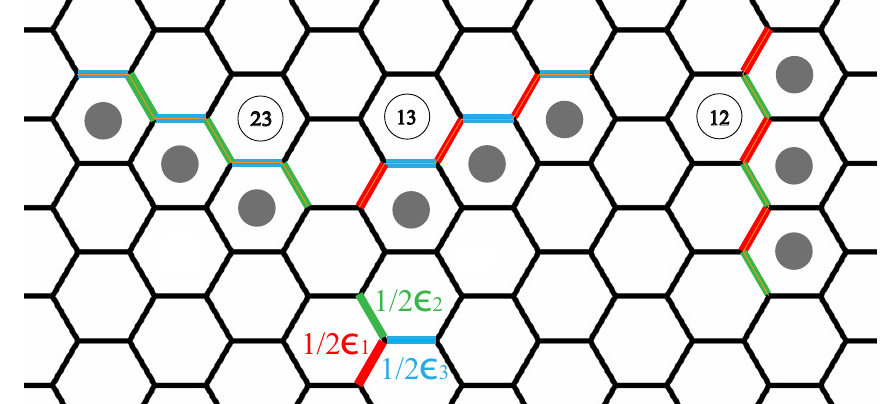}
	\smallskip
	\caption{\small Domain walls at 0 K. Atoms are illustrated as gray circles, again use is made of the dual lattice system. The honeycomb lattice denotes the boundaries between atoms. Color code and segments loads are analogous as the ones in Figure \ref{fig:kink diagram}. Segment of type `{32}' is analogous to segment of type `{23}' and corresponds to the domain wall at 0 K, derived from the expressions for the free energy along the $\hat{\epsilon_3}$ and $\hat{\epsilon_2}$ high-symmetry directions. Segment of type `{13}' corresponds to the domain wall at 0 K, derived from the expression for the free energy along the $\hat{\epsilon_1}$ high-symmetry direction. Segment of type `{12}' is forbidden at 0 K for a freely meandering boundary as the energy per unit length of this segment cannot be found as possible free energy value along any of the directions.}
	\label{fig:boundary diagram}
\end{figure}

To translate the problem into a practical example, it should not matter for $\langle n^2 \rangle$ if we analyze a boundary of the type shown in Figure \ref{fig:boundary diagram} from the main text, say segment `23', from left to right or from right to left. So far we defined all kinks and expressions as if we analyzed from left to right. If we analyzed the boundary from right to left, we should use kink definitions which are going to be directly linked to the expression for $\langle n^2 \rangle$, and most importantly, we should always make sure that the unkinked boundary is made out of 0 kinks, because these will not give contribution to $\langle n^2 \rangle$. So, when counting from right to left, a mirrored (w.r.t. an imaginary vertical line) +1 kink should be labeled as a 0 kink as it is equivalent in energy to a 0 kink when counting from left to right (consult Figure \ref{fig:kink diagram} (B) from the main text), a mirrored +2 kink should be labeled as a -1 kink, a mirrored -1 kink should be labeled as a +2 kink, and so on: positive kinks become negative and with 1 less in magnitude, negative kinks become positive and with 1 more in magnitude.

\begin{figure}[ht]
	\centering
	\includegraphics[width=0.65\linewidth]{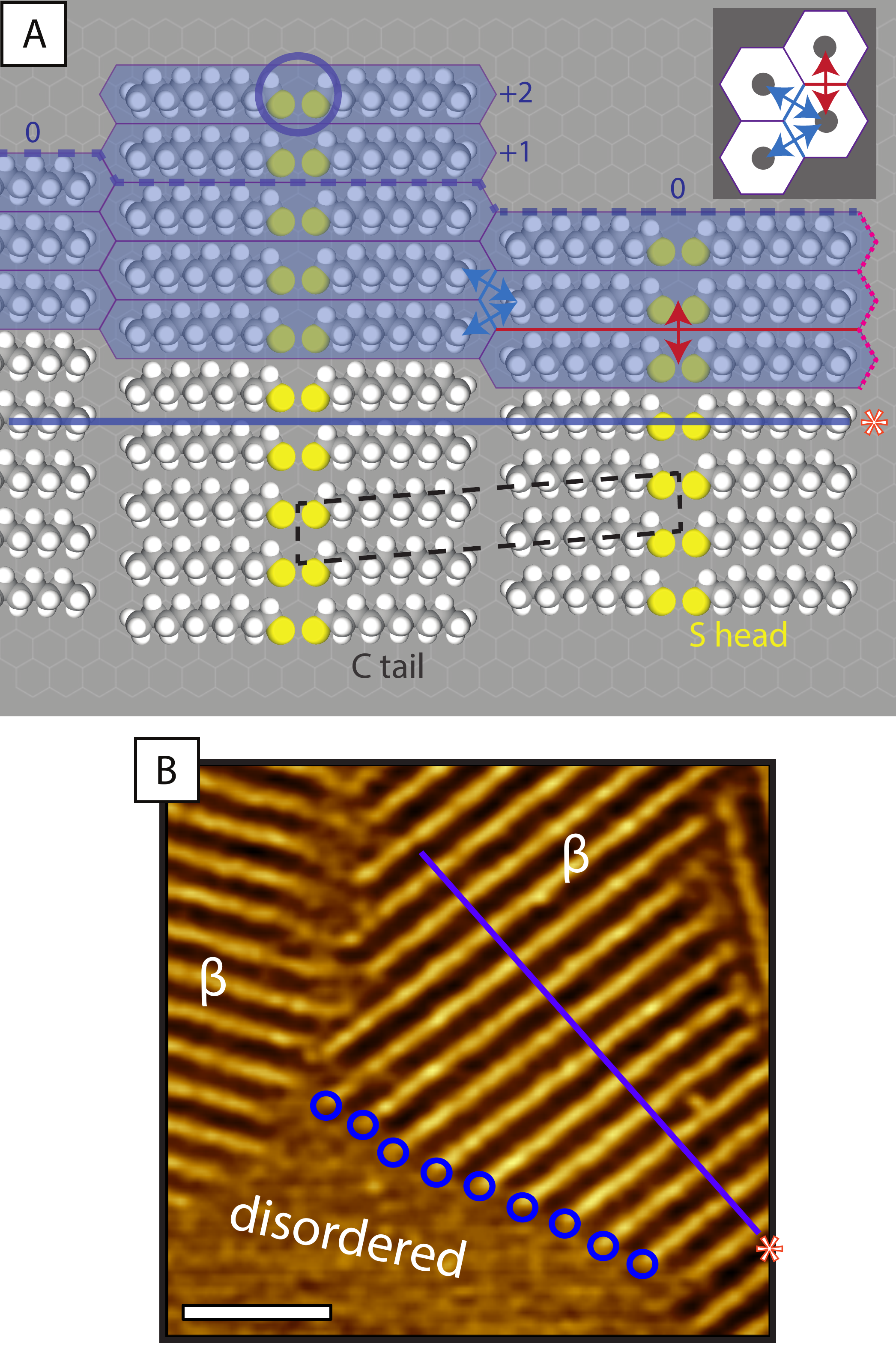}
	\smallskip
	\caption{\small Model of the decanethiol $\beta$ phase (A) and corresponding experimental data (B). (A) The decanethiol molecules on the Au(111) surface (only the honeycomb dual lattice is shown, Au atoms are positioned in the center of each honeycomb cell). A blue horizontal line and a blue circle show the locations corresponding to the same in (B). With a blue dashed line (horizontally drawn) the reference domain wall is shown. With numbers we mark the amount of molecules with respect to the reference domain wall (discussed in the text). Arrows indicate the three interaction energies in the anisotropic triangular lattice (the two in blue are equal, the third, vertically drawn arrow, shown in red is larger). The unit cell (black dashed lines at the lower part of the diagram) in this model contains two molecules. Purple transparent overlay is used analogously to the honeycomb dual lattice (distorted due to the lattice parameters anisotropy). With a vertical pink dashed line on the right the other possible domain wall termination is shown (discussed in the text). In the inset of (A), an illustration of the molecular ordering which neglects the dimensions of the molecules is shown to clarify the choice of the energies (the two molecules per unit cell are shown with a gray circle, the dual lattice is drawn). (B) A typical STM image of the $\beta$ decanethiol phase at room temperature next to a disordered molecular region ($\varepsilon$ phase). The scale bar corresponds to 9 nm. }
	\label{fig:decanethiol}
\end{figure}

As long as the procedures regarding $\langle n^2 \rangle$ described here are followed, there will be no outcome degeneracy problem when selecting different reference orientation along a domain boundary. Note that the sign of the kink can also be omitted as the $\pm n$ kinks will be eventually squared. All these considerations were taken into account when analyzing the decanethiol data in the main text. One can see that based on the current discussion, actually the direction of boundary analysis in Figure \ref{fig:decanethiol} will no longer matter, we would always have the amount of extra/less molecules with respect to the unkinked boundary equal to the kinks. Additionally, then the expression for $\langle n^2 \rangle$ also becomes symmetric because the participating interaction energies are equal.

\newpage
\bibliography{anisotropic_nn}

\begin{thebibliography}{17}
\expandafter\ifx\csname natexlab\endcsname\relax\def\natexlab#1{#1}\fi
\expandafter\ifx\csname bibnamefont\endcsname\relax
  \def\bibnamefont#1{#1}\fi
\expandafter\ifx\csname bibfnamefont\endcsname\relax
  \def\bibfnamefont#1{#1}\fi
\expandafter\ifx\csname citenamefont\endcsname\relax
  \def\citenamefont#1{#1}\fi
\expandafter\ifx\csname url\endcsname\relax
  \def\url#1{\texttt{#1}}\fi
\expandafter\ifx\csname urlprefix\endcsname\relax\def\urlprefix{URL }\fi
\providecommand{\bibinfo}[2]{#2}
\providecommand{\eprint}[2][]{\url{#2}}

\bibitem[{\citenamefont{Bartelt et~al.}(1991)\citenamefont{Bartelt, Einstein,
  and Rottman}}]{Bartelt1991}
\bibinfo{author}{\bibfnamefont{N.~C.} \bibnamefont{Bartelt}},
  \bibinfo{author}{\bibfnamefont{T.~L.} \bibnamefont{Einstein}},
  \bibnamefont{and} \bibinfo{author}{\bibfnamefont{C.}~\bibnamefont{Rottman}},
  \bibinfo{journal}{Phys. Rev. Lett.} \textbf{\bibinfo{volume}{66}},
  \bibinfo{pages}{961} (\bibinfo{year}{1991}).

\bibitem[{\citenamefont{Williams}(1994)}]{Williams1994}
\bibinfo{author}{\bibfnamefont{E.~D.} \bibnamefont{Williams}},
  \bibinfo{journal}{Surf. Sci.} \textbf{\bibinfo{volume}{299--300}},
  \bibinfo{pages}{502} (\bibinfo{year}{1994}).

\bibitem[{\citenamefont{Jeong and Williams}(1999)}]{Jeong1999}
\bibinfo{author}{\bibfnamefont{H.-C.} \bibnamefont{Jeong}} \bibnamefont{and}
  \bibinfo{author}{\bibfnamefont{E.~D.} \bibnamefont{Williams}},
  \bibinfo{journal}{Surf. Sci. Rep.} \textbf{\bibinfo{volume}{34}},
  \bibinfo{pages}{171} (\bibinfo{year}{1999}).

\bibitem[{\citenamefont{Kramers and Wannier}(1941)}]{KramersWannier1941}
\bibinfo{author}{\bibfnamefont{H.~A.} \bibnamefont{Kramers}} \bibnamefont{and}
  \bibinfo{author}{\bibfnamefont{G.~H.} \bibnamefont{Wannier}},
  \bibinfo{journal}{Phys. Rev.} \textbf{\bibinfo{volume}{60}},
  \bibinfo{pages}{252} (\bibinfo{year}{1941}).

\bibitem[{\citenamefont{Onsager}(1944)}]{Onsager1944}
\bibinfo{author}{\bibfnamefont{L.}~\bibnamefont{Onsager}},
  \bibinfo{journal}{Phys. Rev.} \textbf{\bibinfo{volume}{65}},
  \bibinfo{pages}{117} (\bibinfo{year}{1944}).

\bibitem[{\citenamefont{Kaufman}(1949)}]{Kaufman1949}
\bibinfo{author}{\bibfnamefont{B.}~\bibnamefont{Kaufman}},
  \bibinfo{journal}{Phys. Rev.} \textbf{\bibinfo{volume}{76}},
  \bibinfo{pages}{1232} (\bibinfo{year}{1949}).

\bibitem[{\citenamefont{Zandvliet}(2000)}]{Zandvliet2000}
\bibinfo{author}{\bibfnamefont{H.~J.~W.} \bibnamefont{Zandvliet}},
  \bibinfo{journal}{Rev. Mod. Phys.} \textbf{\bibinfo{volume}{72}},
  \bibinfo{pages}{593} (\bibinfo{year}{2000}).

\bibitem[{\citenamefont{Zandvliet}(2015)}]{Zandvliet2015}
\bibinfo{author}{\bibfnamefont{H.~J.~W.} \bibnamefont{Zandvliet}},
  \bibinfo{journal}{Surf. Sci.} \textbf{\bibinfo{volume}{639}},
  \bibinfo{pages}{L1} (\bibinfo{year}{2015}).

\bibitem[{\citenamefont{Zandvliet}(2006)}]{Zandvliet2006}
\bibinfo{author}{\bibfnamefont{H.~J.~W.} \bibnamefont{Zandvliet}},
  \bibinfo{journal}{Europhys. Lett.} \textbf{\bibinfo{volume}{73}},
  \bibinfo{pages}{747} (\bibinfo{year}{2006}), \bibinfo{note}{corrected in:
  Europhys. Lett. {\bf 74}, 1123 (2006).}

\bibitem[{\citenamefont{Sotthewes and Zandvliet}(2013)}]{Kai2013a}
\bibinfo{author}{\bibfnamefont{K.}~\bibnamefont{Sotthewes}} \bibnamefont{and}
  \bibinfo{author}{\bibfnamefont{H.~J.~W.} \bibnamefont{Zandvliet}},
  \bibinfo{journal}{J. Phys.: Cond. Matt.} \textbf{\bibinfo{volume}{25}},
  \bibinfo{pages}{205301} (\bibinfo{year}{2013}), \bibinfo{note}{corrected in:
  J. Phys.: Cond. Matt. {\bf 31}, 499501 (2019).}

\bibitem[{\citenamefont{Sotthewes et~al.}(2013)\citenamefont{Sotthewes, Wu,
  Kumar, Vancso, Sch\"{o}n, and Zandvliet}}]{Kai2013}
\bibinfo{author}{\bibfnamefont{K.}~\bibnamefont{Sotthewes}},
  \bibinfo{author}{\bibfnamefont{H.}~\bibnamefont{Wu}},
  \bibinfo{author}{\bibfnamefont{A.}~\bibnamefont{Kumar}},
  \bibinfo{author}{\bibfnamefont{G.~J.} \bibnamefont{Vancso}},
  \bibinfo{author}{\bibfnamefont{P.~M.} \bibnamefont{Sch\"{o}n}},
  \bibnamefont{and} \bibinfo{author}{\bibfnamefont{H.~J.~W.}
  \bibnamefont{Zandvliet}}, \bibinfo{journal}{Langmuir}
  \textbf{\bibinfo{volume}{29}}, \bibinfo{pages}{3662} (\bibinfo{year}{2013}),
  \bibinfo{note}{corrected in: Langmuir {\bf 35}, 4787 (2019).}

\bibitem[{\citenamefont{Houtappel}(1950)}]{Houtappel1950}
\bibinfo{author}{\bibfnamefont{R.}~\bibnamefont{Houtappel}},
  \bibinfo{journal}{Physica} \textbf{\bibinfo{volume}{16}},
  \bibinfo{pages}{425} (\bibinfo{year}{1950}).

\bibitem[{Bax()}]{Baxter1982}
\bibinfo{note}{Rodney J. Baxter in "Exactly solved models in statistical
  mechanics", Academic Press Limited (London, 1982)}.

\bibitem[{\citenamefont{Wannier}(1945)}]{Wannier1945}
\bibinfo{author}{\bibfnamefont{G.~H.} \bibnamefont{Wannier}},
  \bibinfo{journal}{Rev. Mod. Phys.} \textbf{\bibinfo{volume}{17}},
  \bibinfo{pages}{50} (\bibinfo{year}{1945}).

\bibitem[{\citenamefont{Zandvliet et~al.}(1992)\citenamefont{Zandvliet,
  Elswijk, {van Loenen}, and Dijkkamp}}]{Zandvliet1992}
\bibinfo{author}{\bibfnamefont{H.~J.~W.} \bibnamefont{Zandvliet}},
  \bibinfo{author}{\bibfnamefont{H.}~\bibnamefont{Elswijk}},
  \bibinfo{author}{\bibfnamefont{E.}~\bibnamefont{{van Loenen}}},
  \bibnamefont{and} \bibinfo{author}{\bibfnamefont{D.}~\bibnamefont{Dijkkamp}},
  \bibinfo{journal}{Phys. Rev. B} pp. \bibinfo{pages}{5965--5968}
  (\bibinfo{year}{1992}).

\bibitem[{\citenamefont{Poirier}(1999)}]{Poirier1999}
\bibinfo{author}{\bibfnamefont{G.~E.} \bibnamefont{Poirier}},
  \bibinfo{journal}{Langmuir} \textbf{\bibinfo{volume}{15}},
  \bibinfo{pages}{1167} (\bibinfo{year}{1999}).

\bibitem[{\citenamefont{Poirier et~al.}(2001)\citenamefont{Poirier, Fitts, and
  White}}]{Poirier2001}
\bibinfo{author}{\bibfnamefont{G.~E.} \bibnamefont{Poirier}},
  \bibinfo{author}{\bibfnamefont{W.~P.} \bibnamefont{Fitts}}, \bibnamefont{and}
  \bibinfo{author}{\bibfnamefont{J.~M.} \bibnamefont{White}},
  \bibinfo{journal}{Langmuir} \textbf{\bibinfo{volume}{17}},
  \bibinfo{pages}{1176} (\bibinfo{year}{2001}).

\end{thebibliography}

\newpage

\begin{table}
	\centering
	\smallskip
	\begin{adjustbox}{width=0.6\textwidth}
	\begin{tabular}{c c c c}
		\hline
	    $n_r$ \hspace{2cm} & $n_+$ \hspace{2cm} & $n_-$ \hspace{2cm} & $n_{tot}$ \\
		\hline
		0 \hspace{2cm} & - \hspace{2cm} & - \hspace{2cm} & 266\\
		1 \hspace{2cm} & 76 \hspace{2cm} & 45 \hspace{2cm} & 121\\
		2 \hspace{2cm} & 16 \hspace{2cm} & 10 \hspace{2cm} & 26\\
		3 \hspace{2cm} & 6 \hspace{2cm} & 13 \hspace{2cm} & 19\\
		4 \hspace{2cm} & 7 \hspace{2cm} & 7 \hspace{2cm} & 14\\
		5 \hspace{2cm} & 3 \hspace{2cm} & 1 \hspace{2cm} & 4\\
		6 \hspace{2cm} & 1 \hspace{2cm} & 1 \hspace{2cm} & 2\\
		7 \hspace{2cm} & 2 \hspace{2cm} & 0 \hspace{2cm} & 2\\
		\hline
		total \hspace{2cm} & \hspace{2cm} & \hspace{2cm} & 454\\
		\hline
	\end{tabular}
	\end{adjustbox}
    \smallskip
	\caption{\small Measured number of kinks ($n_{\pm r}$) obtained from analyzing multiple boundaries as the one shown in Figure \ref{fig:decanethiol} (B). $\langle n \rangle^2$ obtained from this data amounts to 1.96.}
	\label{tab:nsquare}
\end{table}

\begin{table}[ht]
	\centering
	\smallskip
	\begin{adjustbox}{width=\textwidth}
		\begin{tabular}{l l l}
			\hline
			$\phi_{\text{start}}\rightarrow\phi_{\text{end}}$ \hspace{2cm} & $\theta(\phi)$ \hspace{2cm} & $\gamma(\theta(\phi))$ \\
			\hline\vspace{0.2 cm}
			$0\rightarrow\pi/6$ \hspace{2cm} & $\phi$ \hspace{2cm} & $[\cos(\theta)-\sqrt{3}\sin(\theta)] \gamma_{\text{AC}}^{0} + 2\sin{\theta}\gamma_{\text{ZZ}}^{\pi/6}$ \\\vspace{0.2 cm}
			$\pi/6\rightarrow\pi/3$ \hspace{2cm} & $\phi-\pi/6$ \hspace{2cm} & $[\cos(\theta)-\sqrt{3}\sin(\theta)] \gamma_{\text{ZZ}}^{\pi/6} + 2\sin{\theta}\gamma_{\text{AC}}^{\pi/3}$ \\\vspace{0.2 cm}
			$\pi/3\rightarrow\pi/2$ \hspace{2cm} & $\phi-\pi/3$ \hspace{2cm} & $[\cos(\theta)-\sqrt{3}\sin(\theta)] \gamma_{\text{AC}}^{\pi/3} + 2\sin{\theta}\gamma_{\text{ZZ}}^{\pi/2}$ \\\vspace{0.2 cm}
			$\pi/2\rightarrow2\pi/3$ \hspace{2cm} & $\phi-\pi/2$ \hspace{2cm} & $[\cos(\theta)-\sqrt{3}\sin(\theta)] \gamma_{\text{ZZ}}^{\pi/2} + 2\sin{\theta}\gamma_{\text{AC}}^{5\pi/3}$ \\\vspace{0.2 cm}
			$2\pi/3\rightarrow5\pi/6$\hspace{2cm} & $\phi-2\pi/3$ \hspace{2cm} & $[\cos(\theta)-\sqrt{3}\sin(\theta)] \gamma_{\text{AC}}^{5\pi/3} + 2\sin{\theta}\gamma_{\text{ZZ}}^{11\pi/6}$ \\\vspace{0.2 cm}
			$5\pi/6\rightarrow\pi$ \hspace{2cm} & $\phi-5\pi/6$ \hspace{2cm} & $[\cos(\theta)-\sqrt{3}\sin(\theta)] \gamma_{\text{ZZ}}^{11\pi/6} + 2\sin{\theta}\gamma_{\text{AC}}^{0}$ \\\vspace{0.2 cm}
			$\pi\rightarrow7\pi/6$ \hspace{2cm} & $\phi-\pi$ \hspace{2cm} & $[\cos(\theta)-\sqrt{3}\sin(\theta)] \gamma_{\text{AC}}^{0} + 2\sin{\theta}\gamma_{\text{ZZ}}^{\pi/6}$ \\\vspace{0.2 cm}
			$7\pi/6\rightarrow4\pi/3$ \hspace{2cm} & $\phi-7\pi/6$ \hspace{2cm} & $[\cos(\theta)-\sqrt{3}\sin(\theta)] \gamma_{\text{ZZ}}^{\pi/6} + 2\sin{\theta}\gamma_{\text{AC}}^{\pi/3}$ \\\vspace{0.2 cm}			
			$4\pi/3\rightarrow3\pi/2$ \hspace{2cm} & $\phi-4\pi/3$ \hspace{2cm} & $[\cos(\theta)-\sqrt{3}\sin(\theta)] \gamma_{\text{AC}}^{\pi/3} + 2\sin{\theta}\gamma_{\text{ZZ}}^{\pi/2}$ \\\vspace{0.2 cm}
			$3\pi/2\rightarrow5\pi/3$ \hspace{2cm} & $\phi-3\pi/2$ \hspace{2cm} & $[\cos(\theta)-\sqrt{3}\sin(\theta)] \gamma_{\text{ZZ}}^{\pi/2} + 2\sin{\theta}\gamma_{\text{AC}}^{5\pi/3}$ \\\vspace{0.2 cm}
			$5\pi/3\rightarrow11\pi/6$ \hspace{2cm} & $\phi-5\pi/3$ \hspace{2cm} & $[\cos(\theta)-\sqrt{3}\sin(\theta)] \gamma_{\text{AC}}^{5\pi/3} + 2\sin{\theta}\gamma_{\text{ZZ}}^{11\pi/6}$ \\\vspace{0.2 cm}
			$11\pi/6\rightarrow2\pi$ \hspace{2cm} & $\phi-11\pi/6$ \hspace{2cm} & $[\cos(\theta)-\sqrt{3}\sin(\theta)] \gamma_{\text{ZZ}}^{11\pi/6} + 2\sin{\theta}\gamma_{\text{AC}}^{0}$\\
			\hline
		\end{tabular}
	\end{adjustbox}
    \smallskip
	\caption{\small The free energy $\gamma(\phi)$ per unit length as plotted in Figure \ref{fig:equil_shape} from the main text. For each portion of the general polar angle $\phi$, the angle $\theta(\phi)$ is given and the formula for $\gamma(\theta(\phi))$ which represents essentially $\gamma(\phi)$. The energies $\gamma_{\text{AC,ZZ}}$ are given in the Appendix text.}
	\label{tab:polar_plot}
\end{table}

\setlength{\footskip}{80pt}

\end{document}